\documentclass[acmlarge, authorversion=true, nonacm=true, natbib=false, screen]{acmart}
\usepackage{fancyhdr}
\usepackage[backend=biber, citestyle=authoryear-comp, uniquename=false, bibstyle=authoryear, isbn=false, maxcitenames=3, maxbibnames=20, dashed=false, backref=true]{biblatex}
\DefineBibliographyStrings{english}{
	backrefpage = {$\to$ p.},
	backrefpages = {$\to$ pp.},
}
\AtEveryBibitem{\clearfield{month}}
\usepackage{hyperref}
\usepackage{microtype}
\usepackage{booktabs}
\usepackage{enumitem}
\setlist{leftmargin=20pt}
\usepackage{caption}
\usepackage{subcaption}
\usepackage{csquotes}
\usepackage{amsmath}
\usepackage{pgfplots}
\pgfplotsset{compat=1.16}
\usepackage{multirow}
\usepackage{booktabs}
\usepackage{tabularx}
\usepackage[nameinlink]{cleveref}

\usepackage{crossreftools}
\usepackage{subcaption}
\usepackage[backgroundcolor=orange!15!white,bordercolor=orange!80!black,textsize=small]{todonotes} 
\tikzset{/tikz/notestyleraw/.append style={rounded corners=0pt,inner sep=0.6ex}}
\presetkeys{todonotes}{inline}{}

\title{Designing Digital Voting Systems for Citizens}
\subtitle{Achieving Fairness and Legitimacy in Participatory Budgeting}

\author{Joshua C. Yang}
\authornote{Corresponding author}
\email{joyang@ethz.ch}
\affiliation{%
  \institution{Computational Social Science, ETH Zürich}
  \country{Switzerland}
}
\author{Carina I. Hausladen}
\affiliation{%
  \institution{Computational Social Science, ETH Zürich}
  \country{Switzerland}
  }
\email{carinah@ethz.ch}

\author{Dominik Peters}
\affiliation{%
  \institution{CNRS, LAMSADE, Université Paris Dauphine - PSL}
  \country{France}}
\email{dominik.peters@lamsade.dauphine.fr}

\author{Evangelos Pournaras}
\affiliation{%
  \institution{University of Leeds}
  \country{UK}}
\email{e.pournaras@leeds.ac.uk}

\author{Regula Hänggli Fricker}
\affiliation{%
  \institution{University of Fribourg}
  \country{Switzerland}}
\email{regula.haenggli@unifr.ch}

\author{Dirk Helbing}
\affiliation{%
  \institution{Computational Social Science, ETH Zürich}
  \country{Switzerland}
}
\email{dhelbing@ethz.ch}

\renewcommand{\shortauthors}{Joshua C. Yang et al.}

\date{\today}

\begin{abstract}
    Participatory Budgeting (PB) has evolved into a key democratic instrument for resource allocation in cities. Enabled by digital platforms, cities now have the opportunity to let citizens directly propose and vote on urban projects, using different voting input and aggregation rules. However, the choices cities make in terms of the rules of their PB have often not been informed by academic studies on voter behaviour and preferences. Therefore, this work presents the results of behavioural experiments where participants were asked to vote in a fictional PB setting. We identified approaches to designing PB voting that minimise cognitive load and enhance the perceived fairness and legitimacy of the digital process from the citizens' perspective. In our study, participants preferred voting input formats that are more expressive (like rankings and distributing points) over simpler formats (like approval voting). Participants also indicated a desire for the budget to be fairly distributed across city districts and project categories. Participants found the Method of Equal Shares voting rule to be fairer than the conventional Greedy voting rule. These findings offer actionable insights for digital governance, contributing to the development of fairer and more transparent digital systems and collective decision-making processes for citizens.
\end{abstract}



\begin{document}

\setcounter{tocdepth}{1}
\newcommand{\nocontentsline}[3]{}
\let\oldaddcontentsline=\addcontentsline
\let\addcontentsline=\nocontentsline
\maketitle
\vspace{0cm}
\hrule
\vspace{-10pt}
\tableofcontents
\vspace{-20pt}
\hrule
\let\addcontentsline=\oldaddcontentsline

\section{Introduction}

Participatory budgeting (PB) is a democratic process that allows citizens to influence government spending in their community. This democratic innovation has emerged as a great success: Hundreds of cities now regularly let their residents vote over how the city government will spend a designated part of its budget \parencite{Wampler2021,DeVries2022}. Originating in Brazil during the 1980s as a process focused on fostering community deliberation, PB has since gained international momentum \parencite{Cabannes2004,Ryan2021}. In its currently predominant form, PB proceeds by asking residents to submit project proposals, and then passes to a formal \emph{voting stage}, where residents vote for the projects they would like to see implemented. The most popular projects are thereby selected for implementation, subject to budget constraints. 

Digitalisation has transformed PB in the recent decades for multiple reasons. Digital platforms can support the scaling of participatory processes and broaden participation, without some of the limitations of purely offline, face-to-face deliberations \parencite{Davies2021AScotland}. Online processes make it easy for cities to configure the platform for their respective use case and to reduce costs \parencite{Palacin2024}. Digital PB, supported by digital means that allow for large-scale participation, also creates space for collective intelligence, which refers to conditions under which a diverse group of people outperform individuals, often even experts \parencite{Page2007}.

The transition to a digitalisation of democratic processes, however, is not without challenges.  Notably, when relying solely on digital tools, the ``digital divide'' \parencite{Cullen2001AddressingDivide} poses a risk to equal access to democratic deliberation. For example, \textcite{LevOn2022} finds significant differences in the budgeting outcomes reached in physical versus remote settings. Also, a large number of projects to be voted on in a PB can make it challenging for voters to take decisions \parencite{Laruelle2021VotingBudgeting}. Furthermore, algorithmic decisions can be perceived as less fair and trustworthy than human decisions \parencite{Lee2018}. Thus, to realise the full potential of collective intelligence, digital democracy systems should be thoughtfully designed from a human-centred perspective, emphasising trust and legitimacy \parencite{Helbing2023DemocracySociety}. 

How do cities currently decide how to set up their digital PB systems? According to a list of PB votes in cities around the world collected by \textcite{Peters2022}, there is much heterogeneity in the rules cities use, and in particular in their choice of the \emph{voting input format}. For example, many cities use a form of \emph{approval voting}, where voters can vote for several projects (but only once per project), typically up to a certain maximum (e.g., up to 3 projects in Rome 2019 and Toulouse 2022, up to 5 in Montreal 2021 and Cambridge, MA, 2016--2021, up to 6 in Grenoble 2019--2022, up to 10 in Lyon 2022, up to 15 in Warsaw 2020--2024).
Other cities give voters a certain number of \emph{points} that they can assign to various projects, and allow them to give several points to the same project (e.g., 5 points in Strasbourg 2019--2021 and Gdańsk 2016--2022, 7 points in Toulouse 2019). Yet other cities allow voters to select a certain number of projects and then \emph{rank} them in order of their preference (e.g., 3 projects in Brest 2022 and Kraków 2019--2021, 5 projects in Gdynia 2016--2019).

However, it is not clear how city officials and legislators decide on these voting input formats and their parameters. There are few academic studies that they can draw on to see which formats are easiest to use, which formats best express voter preferences, and which formats voters prefer to use. The only such study we are aware of is a recent MTurk user study \parencite{Fairstein2023ParticipatoryWorld}, but it has two significant limitations: (1) Several of the input formats it studied are popular among voting theorists but not used by cities (e.g., ranking all 20 projects, or distributing 100 points across projects), and therefore the results aren't directly informative for the choices that city officials make. (2) \textcite{Fairstein2023ParticipatoryWorld} used a between-group design, where each participant only voted using a single voting format. As a consequence, they  could not ask participants for their preferences among input formats. 

We asked 180 participants to vote in a fictional PB program. Participants were drawn from the joint subject pool of students of the two major universities in Zurich (ETH Zurich and the University of Zurich).  The fictional PB program involved 24 projects, whose descriptions we adapted from proposals of Zurich's 2022 PB (``\href{https://mitwirken.stadt-zuerich.ch/processes/stadtidee}{Stadtidee}''). Via an online interface, we let participants vote over the same list of projects 6 times in a row, using a different input format each time. We chose to study the following formats: selecting any number of projects, selecting 5 projects, distributing 5 points across projects, distributing 10 points, ranking 5 projects, and distributing 10 points across 5 chosen projects. After the vote, we asked participants how easy they found each format to use (approval and ranking were rated easiest) and how well it captured their preferences (ranking and distributing 10 points were rated best). We also asked participants which input format they would recommend a city to use for their PB, by giving a preference ordering of the input formats that they had just used. Ranking 5 projects was a clear favourite.

The choice of input format is only the first decision that needs to be made to organise a PB vote. The second decision is which voting aggregation rule to use, i.e., the rule that, given the votes, decides which projects are winning \parencite{Rey2023}. At this point, most cities around the world use a simple aggregation method that follows a majority rule to select projects according to the number of votes they accumulated, which is known as the \emph{Greedy rule} in the computer science literature. This rule designates the projects with the highest vote count as winners, until the budget runs out \parencite{Talmon2019}. Recent literature has identified important downsides of this rule \parencite{Peters2021,Faliszewski2023,Boehmer2023}: 
\begin{itemize}[leftmargin=20pt]
	\item The rule is \emph{majoritarian}: a plurality of voters can monopolise the entire budget, while the interests of smaller groups may be ignored.
	\item The rule tends to favour projects in the city centre and to neglect less densely populated areas.
	\item The rule has low sensitivity to changes in project costs, leading to strategic bundling and cost inflation.
\end{itemize}
Researchers have developed an alternative voting rule known as the \emph{Method of Equal Shares} (MES) \parencite{Peters2021} that supports \emph{proportional representation} in PB and thereby creates fair opportunities for small groups and projects to win \parencite{Aziz2017ProportionallyAlgorithms}. 
MES works by assigning each voter an equal share of the available budget, and then only selects a project as a winner if the supporters of that project together have enough budget to cover the project's costs.
This approach strives for every voter to have an equal influence on the voting outcome, whenever possible.
In 2023, MES has been implemented for the first time for PBs in the cities of Aarau (Switzerland) and Wieliczka and Świecie (Poland).

While there are well-developed \emph{theoretical} arguments for using MES (in particular, formal fairness guarantees, see, e.g., \cite{Peters2021,Brill2023}), and while the implementations in Switzerland and Poland appear to have been successes,%
\footnote{See, for example, news media coverage (Rzeczpospolita, 2023-04-28, \href{https://www.rp.pl/polityka/art38409241-czy-czeka-nas-rewolucja-w-budzetach-obywatelskich}{\textit{Czy czeka nas rewolucja w budżetach obywatelskich?}} (Are we facing a revolution in participatory budgeting?); ArgoviaToday, 2023-07-05, \href{https://www.argoviatoday.ch/aargau-solothurn/aarau-olten/das-passiert-mit-den-50000-franken-in-aarau-152334880}{\textit{Das passiert mit den 50'000 Franken in Aarau}} (What happens with the CHF 50,000 in Aarau)), or the analysis of \textcite{Boehmer2023}.}
there are no studies on the questions of which voting rule is preferred by voters and whether voters feel that fairness in the sense of proportional representation is a desirable feature of a voting rule for PB.

To answer these questions, in a second part of our study, we presented participants with a fictional outcome of a PB program, involving the same 24 projects they had seen in the first part of the study. Two outcomes were displayed, one obtained by ``Method A'' and the other by ``Method B''. (These outcomes correspond to the outcome of the Greedy rule and the Method of Equal Shares, respectively.) We asked participants how satisfied they were with each outcome and how fair they perceived them to be. We then divided the participants into three groups to see how their views would change in response to different types of \emph{explanations} of the outcomes they saw, mirroring ways in which cities might choose to communicate the voting results. 
\begin{itemize}[leftmargin=20pt]
        \item The first group was given a detailed explanation of how the two methods work, explaining how Method A selects the projects with the most votes, while Method B equally distributes the available budget.
	\item The second group was shown graphs with statistical information about the distribution of voter utility. For example, they saw that, for Method A, voters on average approved a higher fraction of the spending. They also saw that, under Method B, a larger share of voters got at least one of the projects which they voted for, funded.
	\item The third group was shown graphs with statistical information about how the two methods distributed the budget across city districts and project categories (transportation, culture, or nature), and how the relevant percentages compared to the vote shares of the categories and districts.
\end{itemize}
After going through the explanation, we asked participants again what they thought of the two methods. Their answers indicated that they thought Method B (i.e., MES) provided fairer outcomes.

We hope that our study provides action-guiding information for governments when deciding how to design their PB digital voting platform, how to choose the voting rule, and how to communicate about the voting outcome. We also believe that our results are of interest to the subject of digital democracy more generally, as they underline the value of asking participants for their preferences about the way democratic institutions should be structured.

\section{Related Work}

In the literature review below, we examine existing research on digital PB, particularly in relation to issues of trust, fairness, and legitimacy, as well as relevant research from Computational Social Choice, with a focus on prior behavioural experiments regarding PB.

\subsection{Evolution and Challenges of Digital Participatory Budgeting}

PB is a democratic process where community members directly decide how to spend a portion of public budget. It allows citizens to identify, discuss, and prioritise public projects, and gives them the power to make real decisions about how to spend money \parencite{Wampler2000}. A significant evolution in PB over the past decade has been the shift towards online participation. As \textcite{Wampler2021} point out, efforts to include greater participation and overcome the high costs associated with large-scale face-to-face participation are at the core of digital PB programs. In cities like Helsinki, Paris, Madrid, and Barcelona, the adoption of digital platforms enables the collection of a much larger volume of votes and facilitates the use of more sophisticated voting inputs and/or aggregation methods.

However, the digitalisation of PB is not without challenges. A primary concern is the ``Digital Divide'', as noted by \textcite{Cullen2001AddressingDivide}, which poses a risk to equal access and participation in PB for all citizens due to the difference in their digital skills. Concerns about trust in online voting, vote manipulation, and vote security are also paramount. These challenges underscore the importance of adopting inclusive strategies and robust security measures in digital PB to support participation and trust in the process. 

Note that PB is a form of ``multi-candidate election''. In electoral processes that involve voters selecting from more than two options~\parencite{Merrill1984}, a common challenge is managing ``information overload'', which can significantly impact democratic decision-making processes \parencite{Roetzel2019InformationDevelopment}. A connected idea is the ``Choice Overload'' dilemma, where an excess of options can overwhelm voters, as explored by \textcite{Iyengar2000WhenThing} and \textcite{Laruelle2021VotingBudgeting}. An illustrative case is Tartu's PB program, which demonstrated that limiting project choices can enhance voter participation. Eventually, this led to a city-imposed cap of 25 projects \parencite{Mre2021IncreasingProcesses}.

\subsection{Trust, Fairness, and Legitimacy in the Digital Age}
\label{sec:trust-legitimacy}

As the digital era advances, trust and legitimacy emerge as pivotal elements in the political landscape, especially in the context of decision-making processes and the digital transformation of cities. \textcite{Helbing2023DemocracySociety} emphasises the role of digital technologies in promoting participatory democracy, stressing the importance of accessibility and fairness to avoid reinforcing existing power imbalances.

\emph{Political trust} refers to the confidence that stakeholders have in institutions, specifically with regard to acting competently and ethically \parencite{Uslaner2002}.
\textit{Fairness} as a term comprising many meanings. In the PB context, \textit{distributive fairness} and \textit{proportional fairness} can achieve that resource allocation proportionally reflects collective community preferences, where no subgroup is able to achieve a more favourable outcome by reallocating their budget share \parencite{Fain2016}. 

In this study, we pay particular attention to \textit{perceived fairness}. This means a procedure or outcome that is ``perceived by individuals or collectives as fair according to previous norms or standards'' \parencite{Peiro2014}. In this way, our findings can be realistically applied to real-world social settings and connected to the idea of political trust and legitimacy.

\textit{Legitimacy} is the perceived acceptability of an institution's actions, often gauged by their adherence to democratic principles, policy outcomes, trust, and procedural fairness \parencite{Beetham1991, Scharpf1999}. 
\emph{Output legitimacy} reflects citizens' acceptance of the results of their government's policies and laws, which faces various challenges in the digital age. For example, the survey by \textcite{hartley2021} reveals a dichotomy: while citizens appear to appreciate the benefits of smart cities, they tend to distrust their privacy and security measures, overall feeling dissatisfied with their participation in policy-making. This underscores the challenge of balancing democratic procedures against policy outcomes \parencite{Strebel2019TheCountries}, i.e., of balancing input and output legitimacy \parencite{Scharpf1998InterdependenceLegitimation}. Altogether, this calls for the concept of ``throughput legitimacy'', which evaluates the efficiency and quality of the processes that transform inputs into outputs \parencite{Schmidt2010DemocracyPaper}.

In the digital governance context, \textcite{Hanggli2021Human-centeredElements} advocate a human-centred approach, which marries inclusive voting mechanisms with the integration of diverse opinions and the synergy between digital tools and real-life interactions. The human-centric approach does not only aim to bolster legitimacy and trust in governance, but also emphasises the critical need for further research into democratic innovations that enhance citizen participation in the digital age.

Another relevant concept is \emph{Explainable AI} (XAI). As algorithmic decision-making becomes commonplace in people's everyday lives, XAI has emerged as a crucial approach to address the issue of trust in algorithmic ``black boxes''. In discussions on XAI, a fundamental distinction exists between ``explanation proper'' and ``explanation as justification'' \parencite{Kenny2021, Sormo2005} The ``explanation proper'' (also known as \emph{pre-hoc explanation} or \emph{model transparency}) focuses on explaining how an AI model works. This type of explanation aims to make the internal mechanisms and decision-making processes of the model inherently understandable. On the other hand, ``explanation as justification'' (or \emph{post-hoc explanation}) seeks to justify or explain after the fact why the system produced a given answer. This approach does not require the model itself to be transparent, but instead uses various techniques to interpret and explain the model's outputs.

The distinction between pre-hoc and post-hoc explanations reflects a trade-off between accuracy and comprehensibility in XAI. Pre-hoc explanations prioritise inherent understandability. However, requiring that such explanations are available might limit the complexity and performance of algorithms. Post-hoc explanations focus on making outcomes easily understandable. However, such explanations might be too simple to fully capture the model's decision-making process. We will apply these XAI approaches to our study of voting in PB, and compare 3 explanation types of the voting process and outcome. 

\subsection{Computational Social Choice Literature on PB}

Researchers in Computational Social Choice study voting from a computational perspective \parencite{Brandt2016}. Participatory budgeting is interesting from this point of view, as the selection of a budget-feasible set of winning projects is a combinatorial collective decision problem. In particular, the approach belongs to the class of multi-winner voting problems \parencite{Faliszewski2017}, which have received much attention in recent years. For an overview of recent work on participatory budgeting, see the survey by \textcite{Rey2023}.

The seminal paper of \textcite{Benade2021} compared different voting input formats in the framework of \emph{implicit utilitarian voting}, where the goal was to find a winning set that maximises utilitarian social welfare despite eliciting only partial information about voter preferences. \textcite{Benade2021} proved that a certain variant of approval voting (randomised threshold approval voting) allowed for good outcomes on their worst-case measure. They also studied \textit{knapsack voting} (where voters were asked to select a highest-value set of projects that fits in the budget constraint \parencite{Goel2019}), which performed much worse, as well as some input formats based on complete rankings of projects. However, 
the voting rules implicitly proposed in their work would be difficult to explain and justify to voters. In addition, the input formats studied are different from the ones that are common in practical use.

Most other work in the Computational Social Choice literature focuses on the design and evaluation of different voting rules, i.e., methods of aggregating votes and deciding on sets of winning projects. Researchers have studied the properties of the standard Greedy method and its variants \parencite{Talmon2019,Baumeister2020,Los2022} as well as related rules that maximise social welfare objectives \parencite{Hershkowitz2021,Laruelle2021VotingBudgeting,Fluschnik2019}.

An issue with the Greedy method and related approaches is that they are majoritarian and may neglect minority interests, which motivates the search for a PB voting rule that provides \emph{proportional representation} \parencite{Aziz2017ProportionallyAlgorithms}. \textcite{Peters2021} proposed the \emph{Method of Equal Shares} (MES), a simple voting rule based on the idea of dividing the available budget equally between the voters and then electing popular projects when their supporters have sufficient remaining budget to fund the projects. They proved that MES provides a proportional representation guarantee called EJR (``extended justified representation'', \cite{Aziz2017JR}). Several other papers established additional formal proportionality properties of MES \parencite{Brill2023,Brill2023a,Los2022}. \textcite{Faliszewski2023} simulated MES on datasets from real-world PB and found that it produces fairer outcomes than the Greedy method. MES has since been implemented in PB voting in Aarau, Wieliczka, and \'{S}wiecie (see \href{https://equalshares.net/elections}{equalshares.net/elections}).

\subsection{Prior Experimental Work}

The work on voting in PB we mentioned above has only used theoretical and simulation approaches. Overall, there has been little work testing these ideas with human subjects. Below, we discuss three relevant studies.

In an unpublished study with 1,200 subjects from Amazon Mechanical Turk, \textcite{Benade2018} evaluated the efficiency and usability of different input formats in PB. They found that $k$-approval voting imposes a low cognitive burden on voters, although the participants themselves did not always perceive this to be the case. The study was framed as the participants being stranded on a desert island -- what items they would wish to have in such a situation. Therefore, the applicability to PB in cities is unclear. In addition, some of the input formats studied were far removed from those used in practice.

\textcite{Fairstein2023ParticipatoryWorld} recruited more than 1,800 subjects from Amazon Mechanical Turk. In a similar setup to our study, they had each participant enter a vote in a fictional PB. Some participants saw a PB with 10 projects, others saw 20 projects. Each participant voted only once, using one of six input formats: Select up to 5 projects, select projects up to the budget limit (``knapsack vote''), rank all projects, rank all projects in order of ``value for money'', distribute 100 points between projects, and select all projects you would give more than 10 points if you were to divide 100 points (``threshold approval''). The voting interface displayed a category symbol for each project, as well as a fictional city map showing project locations. The authors measured the time taken by participants, and asked participants to answer the following questions on a 1--5 scale: ``How easy did you find the voting task?'' and ``How well did the input format capture your preferences?''. They found that selecting up to 5 projects was the input format that took the least time. It also scored highest on both rating questions. Participants further rated the points input format highly, even though it took more time to use. Our study differs from this by letting the same participants vote several times, using different input formats, so that participants can compare them. We also used different input formats that are closer to the ones in practical use.

The two studies discussed above focused on the choice of the input format. Regarding the choice of the voting rule (i.e., how to aggregate the votes into a set of winning projects), in an unpublished paper, \textcite{Rosenfeld2021} present a study with 215 subjects aimed at identifying what non-experts deem as ``fair'' or ``desirable'' outcomes in PB programs. In particular, they showed participants a small vote matrix of how 5 voters assigned utilities to 5 projects, and then asked which set of projects they would choose to fund, if placed in the position of social planner. They also asked participants to express their opinions on statements such as ``Every voter should get at least one of her approved projects funded (if possible)'' or ``The most approved projects should be funded''. However, this study did not feature the two voting rules that we focus on (Greedy and MES).

Finally, note that a domain related to participatory budgeting is that of \emph{threshold public goods}, where a cost threshold must be met for community-wide benefits to be realised. These have been a focus of studies in behavioural economics \parencite{Croson2000, Corazzini2015}.

\section{Formal Model}

To understand the setup of our study, we need to formally define the two voting rules we are interested in -- Greedy and MES -- and how they interact with different voting input formats. We are working in a setting where we have discrete projects that can either be fully funded or not funded at all, and we are subject to a budget bound. This is the model of \emph{bounded discrete PB} \parencite{aziz2021participatory}. Let \( P = \{p_1, \ldots, p_m\} \) be the set of available projects, and let \( B \) be the budget limit. Each project $p \in P$ has a cost $c_p > 0$. A set $W \subseteq P$ of projects is \emph{feasible} if $\sum_{p \in W} c_p \le B$. Our goal is to select a feasible set of winning (i.e., funded) projects based on \emph{votes}.

Let $N = \{1, \dots, n\}$ be the set of voters. These voters submit their preferences using a \emph{voting input format}, and we will translate their submissions into an \emph{additive utility function}, so that each voter $i \in N$ has their preferences represented by a function $u_i : P \to \mathbb{R}_{\ge 0}$. Here, $u_i(p)$ reflects the (imputed) utility that voter $i$ obtains from project $p$ being selected. Then $u_i(W) := \sum_{p \in W} u_i(p)$ is the (imputed) overall utility that voter $i$ enjoys when $W$ is the set of winning projects.

There are two main schemes for translating voting inputs into utility functions that have been discussed in the literature, known as \emph{cost utilities} and \emph{cardinality utilities} \parencite{Brill2023,Rey2023}. Intuitively, when using approval voting, under cost utilities a voter's satisfaction is measured as the \emph{total amount of funding} that went to approved projects previously voted for, while under cardinality utilities it is the \emph{number} of approved projects that were selected for funding.
Formally, the cost utility translation scheme defines utilities as follows, for the contexts of approval voting and of point voting, respectively:
\begin{align*}
u_i(p) = 
\begin{cases}
	c_p & \text{if $i$ approves $p$} \\
	0 & \text{otherwise} \\
\end{cases}
\qquad
&\text{and}
\qquad
u_i(p) = 
\begin{cases}
	c_p \cdot s_i(p) & \text{if $i$ assigns $s_i(p)$ points to $p$} \\
	0 & \text{if $i$ assigns no points to $p$} \\
\end{cases}
\intertext{The cardinality utility translation scheme, in contrast, defines utilities as follows:}
u_i(p) = 
\begin{cases}
	1 & \text{if $i$ approves $p$} \\
	0 & \text{otherwise} \\
\end{cases}
\qquad
&\text{and}
\qquad
u_i(p) = 
\begin{cases}
	s_i(p) & \text{if $i$ assigns $s_i(p)$ points to $p$} \\
	0 & \text{if $i$ assigns no points to $p$} \\
\end{cases}
\end{align*}
Note that we treat ranking input as a type of point voting, by converting ranks into points. Specifically, in one of our input formats (``S5R''), we ask participants to rank 5 projects, telling them before this means that 5 points will be assigned to their top choice, 4 points to their second choice, and so on.

We can now formally define the two voting rules that we study in this paper.

\paragraph{Greedy Rule} 
The Greedy rule sorts the projects in order of decreasing ratio of total utility to cost, i.e., $\sum_{i \in N} u_i(p)/c_p$. It starts with an empty set $W = \emptyset$ of winners, and goes through the sorted list project by project, starting with the highest-valued project. For each project $p$, it checks whether it can be added to $W$ without violating the budget constraints, i.e., if $\sum_{p'\in W} c_{p'} + c_p \le B$. If so, it adds $p$ to $W$.

\paragraph{Method of Equal Shares (MES)} 
MES assigns each voter $i \in N$ an equal budget $b_i \ge 0$ (the value of which will be determined later). 
For $q \ge 0$, we say that a project is $q$-affordable if its cost can be funded by voters paying at most $q$ monetary units per unit of utility, i.e., if
\[ c_p = \textstyle\sum_{i \in N} \min(q \cdot u_i(p), b_i). \]
The rule starts with an empty set $W = \emptyset$ of winners. It then checks if any remaining project is $q$-affordable for any $q$. If no project is affordable, the rule finishes and returns $W$. Otherwise, it selects a project $p$ that is $q$-affordable with minimum $q$ and adds it to $W$. It also reduces voters' budgets to account for the cost of $p$ by setting $b_i \gets b_i - \min(q \cdot u_i(p), b_i)$ for each $i \in N$. The value of the starting budget is determined by starting with a starting budget of $B/n$, running the rule, and repeatedly increasing the starting budget by 1 unit if this is possible, i.e., if the cost of the selected set $W$ does not overshoot the budget limit. This procedure of budget increase is known as ``Add1'' in the literature \parencite{Faliszewski2023}.

\medskip
In our formal setup, the rule used by most cities running PB -- select the projects that received the highest total number of points -- is equivalent to the Greedy rule with cost utilities. (Note that cost utilities multiply the number of points by the project cost, but then the definition of the Greedy rule divides that value by the project cost.) Recently, some cities have started using MES with cost utilities. Because the use of cost utilities is standard in implementations, when we refer to Greedy and MES, we implicitly mean their cost-utility versions. We refer to the cardinality-utility versions as \emph{Economical Greedy} and \emph{Economical MES} because these rules proceed in order of vote count divided by project cost. 

While studying voting input formats, we will explore the effect of adopting a format on the output according to each of the four aggregation methods (Greedy and MES, with cost utilities and with cardinality utilities). While studying participants' rating of outcomes and how it changes when presenting different explanations, we will focus on the outcomes obtained by the cost-utility variants of Greedy and MES (since these are the standard implementations).

\section{Experimental Design}
\label{sec:design}

We performed a study in which participants could vote in several ways in a fictional PB program and express their views and preferences.
The experiment was pre-registered on AEA ACT Registry,\footnote{https://www.socialscienceregistry.org/trials/11021} approved by the ETH Zurich Ethics Commission (approval number: EK 2022-N-143) and conducted in March 2023 via the online platform Qualtrics in English language. All participants provided informed consent. 

A total of 180 participants were sourced from the joint subject pool of the two major universities in Zurich: ETH Zurich and the University of Zurich. To be eligible for participation, individuals had to be registered students at one of these universities. We could therefore assume that participants were overall familiar with the city of Zurich. (In response to the question ``to what extent do you feel connected with the city of Zürich?'', 108 participants said ``a lot'' or ``a great deal'', 52 said ``a moderate amount'', 17 said ``a little'', 3 said ``not at all''). Our general impression was that the participants were very engaged during the study. Many participants volunteered detailed thoughts in free text fields, where we invited them to explain their responses.

The study started with several screens giving a high-level explanation of what participatory budgeting is about and how it works. When asked if participants knew about PB before today, 16 answered ``yes'' and 164 answered ``no''. We then familiarised participants with the idea of voting in PB by going through a brief tutorial. In this test run, the participants were situated in a hypothetical city with 3 different districts and exposed to the concept of different voting inputs and voting aggregation methods. 

The main task of the participants was to collectively decide on allocating 60,000 CHF to various projects within the city of Zurich, as part of a fictional PB program. 
Participants were made aware that their payoff in the experiment would be linked to their gains in the simulated PB, thus providing a tangible incentive for their decision-making within the study. The description was: ``\textit{On top of the show-up fee, you can win an extra 0-20 CHF in this democratic game. The higher the amount of budget you can win for your selection of projects in the collective outcome determined by all 180-200 participants, the more you will be paid in this experiment.}'' Note that this induces cost utilities, though we did not observe systematically higher vote counts for the more expensive projects. With the variations in payoffs (CHF 20), we chose this approach to encourage the participants to take the task seriously.

To mirror a realistic PB setting, where voters have real preferences for urban projects, participants were asked to identify the city district in Zurich that they ``feel connected to the most'' (Nord [North], Ost [East], Süd [South], West), and to indicate how much they care about different categories of urban projects (Transportation and traffic organisation; Culture; Nature and urban green areas), by dividing 100 points between the three categories according to their intensity of preference. (For simplicity of analysis, we identified participants with the category they assigned the highest rating to.) This initial part was designed to situate the voters in the city and prompt them to think about realistic urban concerns in order to simulate a real PB setting.

Participants were then shown a list of 24 projects and asked to think about their preferences. The 24 projects are shown in \Cref{fig:random-outcome-table}. The majority of project descriptions was adapted from the PB ``Stadtidee'' that Zurich ran in 2022, with some additional projects created to have an equal number of projects in all 4 districts, all 3 categories, and all 2 different cost levels. The projects were presented in a table, showing a project's ID, its name, its district, its category, and its cost. For simplicity, each project was assigned a cost of either CHF 5,000 or CHF 10,000. Projects were displayed sorted by district and then by category. For each district-category combination, there were exactly 2 projects (a cheap and an expensive one), so that overall there were 6 projects per districts and 8 projects per category. Projects were shown in the same order across all screens to reduce participant workload.

The experiment was designed to answer two research questions, regarding the choice of voting input format and the choice of aggregation rule, respectively.

\subsection{Research Question 1: What are voters' preferred voting input formats, and how does the choice of voting input formats impact the collective outcome?}
\label{sec:design:inputs}

Once participants had reviewed and understood the 24 project descriptions on a preview page, participants proceeded to cast their votes, using six voting input formats for PB. We chose these formats for being similar to those used by cities in practice. \Cref{fig:input-screenshots} in the Appendix shows screenshots of the different interfaces.

\begin{itemize}
	\item SN: \emph{Select any number of projects}. This is the standard approval voting \parencite{Brams1978}, without any constraints imposed, except that we required the selection of at least one project. The prompt was: ``You can select all the projects you approve of. Your vote will be perfectly valid no matter if you vote for many or for few projects.''
	\item S5: \emph{Select 5 projects}. This is 5-approval voting, mirroring a commonly used input format. We decided to require participants to select exactly 5 projects, to make responses better comparable.
	\item D5: \emph{Distribute 5 points}. This screen displayed a slider next to each project, allowing participants to select a value between 0 and 5 for each project (with sliders starting at 0). The system enforced the chosen values to sum up to 5. The prompt was: ``In this vote, you distribute 5 points among the projects you like. You can concentrate your points on one project, or spread them across a few projects.''
	\item D10: \emph{Distribute 10 points}. Using the same setup as for D5, but with sliders allowing values between 0 and 10, enforcing a total of 10 points assigned.
	\item S5R: \emph{Select 5 and rank}. This screen had a drag-and-drop interface allowing participants to rank up to 5 projects. The prompt was: ``In this vote, you first drag the top 5 projects that you like to the box, then rank them in order. The top project will get 5 points; the second project, 4 points; the third, 3 points; and so on.''
	\item S5D10: \emph{Distribute 10 points across 5 projects}. This screen showed the participants only the 5 projects they had ranked in S5R and asked them to distribute 10 points across them using the same interface as for D10. We re-used the selection of 5 projects from the prior step to reduce participant workload.
\end{itemize}

Note that all participants used the input formats in the same order, so that their experience followed a logical sequence to avoid confusion. However, when interpreting results, we need to keep in mind that participants may become more used to expressing their preferences for later input formats.

After submitting their votes, participants evaluated each of the input formats with regard to several criteria. 
\begin{itemize}
	\item \emph{Easiness}: ``How difficult do you find using each of the voting input methods?'' (5-point scale from ``extremely easy'' to ``extremely difficult'').
	\item \emph{Expressiveness}: ``How well does each voting input method capture your actual preferences for these projects?'' (5-point scale from ``does not describe my preference'' to ``clearly describes my preference'').
	\item \emph{Recommendation}: ``What voting input method would you recommend a city to use in a Participatory Budgeting program?'' (by ranking the six input formats in the order of preference).
\end{itemize}

We then asked participants to explain the reasons for their evaluations, in particular the reasons why they chose to rank an input format in the first or  last place with regard to the Recommendation question.

Finally, we asked participants to report how important various project characteristics were to them when deciding which projects to vote for. The characteristics were ``the district/location'', ``the categories/topics'', ``the cost of the project'', and ``how likely they can win''. For each characteristic, we provided a five-point scale from ``not important at all'' to ``extremely important''.

\subsection{Research Question 2: How do voter perceptions of fairness and trustworthiness vary across aggregation methods such as Greedy and MES, and what effect do different explanations have on these perceptions?}

In order to investigate voter perception of different voting outcomes, participants were shown simulated outcomes which were calculated using the two aggregation methods Greedy and MES. These outcomes concerned the same PB instance with 24 projects that participants were familiar with after their votes.

We generated two simulated voting instances using a random procedure, each based on the votes of 200 fictitious voting agents using the S5 and D5 voting input format, respectively.
\begin{itemize}
	\item Instance 1 was obtained by having agents select a set of 5 projects uniformly at random from all 5-subsets of the 24 available projects.
	\item Instance 2 was designed to be ``polarised'', with agents preferring certain projects. For each agent, we assigned them randomly to one of the 4 districts and randomly to one of the 3 categories (independently, in each case uniformly at random). We then identified the 2 projects matching the agent's assigned district-category preference, and gave them a weight of 6, with the remaining 22 projects getting a weight of 1. We then randomly sampled a project 5 times (i.i.d., with probabilities proportional to the weights, with replacement), giving a point to a project when sampled. Note that the weights translated to an expected 1.76 out of 5 points assigned to focus projects.
\end{itemize}
For each instance, we then converted the resulting votes in D5 format to utilities, using the cost utility translation scheme, and we used the Greedy rule and MES to compute the sets of winning outcomes. As expected, the two rules produced relatively similar outcomes in Instance 1, but quite distinct outcomes in Instance 2, due to its polarised nature (see \Cref{fig:random-outcome-table,fig:polarised-outcome-table}).
We divided participants into two groups, each seeing one of the two instances in this part of the experiment. Participants were shown a result table that displayed the outcomes, labelled A and B (corresponding to Greedy and MES, respectively, but participants were not told these names). This table also highlighted the districts and categories of the projects, allowing participants a comparative view. These outcome tables are shown in \Cref{fig:random-outcome-table,fig:polarised-outcome-table} in the Appendix.

We instructed participants that these results were simulated (and therefore in particular not based on their earlier votes). We then asked participants to rate the outcomes:
\begin{itemize}
	\item ``Considering how you voted, how satisfied are you with these two different outcomes?'' (5-point scale from ``extremely dissatisfied'' to ``extremely satisfied''),
	\item ``How fair do you perceive the two outcomes?'' (5-point scale from ``extremely unfair'' to ``extremely fair''),
	\item ``Say outcome A and B were produced using aggregation method A and B, respectively. Based on your intuition, how trustworthy do you perceive these two aggregation methods?'' (5-point scale from ``strongly disagree'' to ``strongly agree'').
\end{itemize}
We also asked participants to explain the reasons for their answers.

One might be hesitant to adopt voting rules such as MES because they are harder to explain to voters than the straightforward Greedy rule. Thus, we explored how participants would react to different ways of explaining the voting outcomes.
We divided participants into three groups, and each group was shown a different type of explanation. Inspired by different approaches in XAI, we designed three explanations, which consisted either of a description of the methods used to compute the winning sets of projects, or of statistics illustrating the resulting voting outcomes.
\begin{itemize}
	\item \emph{Mechanism explanation}: 
	Connected to the XAI idea of model transparency, participants in this group received a walk-through of how the two voting rules work, i.e., of the mechanism by which the Greedy and MES outcomes were computed. For each method, we displayed an about 100-word description defining the method, together with a simple infographic (for Method A, showing a list of projects sorted by vote count; for Method B, showing a diagram explaining how the budget flows from city to voters and from voters to projects, similar to a graphic used on the website \href{https://equalshares.net/}{equalshares.net}).  
	\item \emph{Individual utility distribution explanation}: In this group, we showed participants a graph of the distribution of cost utilities produced by the Greedy and MES outcomes for the 200 simulated voters. Adapting the notion of post-hoc explanation in XAI, the graph showed monetary amounts on the $x$-axis, with each of the two outcomes inducing a line that, for each monetary amount, indicated the number of simulated voters who approved winning projects, whose cost summed up to that amount. Inspecting these graphs, it was clear that Outcome A left a larger number of voters with utility 0, while also giving some voters very high utility; in contrast, Outcome B reduced the number of voters at both extremes and ensured that most voters had a moderate amount of utility. This reflects the behaviour of the two voting rules when simulated on real PB data \parencite{Faliszewski2023}. 
	\item \emph{Category and district explanation}: Participants in this group were shown how the two outcomes divide the budget between projects from different districts and different categories. For districts, they were shown three pie charts in a row, the first one showing the division of the population of 200 simulated voters between the 4 districts; the second one showing the proportions of budget spent by Outcome A in the districts, and the third one showing this for Outcome B. For categories, they were shown a similar row of three pie charts. Consistent with findings from simulations on real PB data \parencite{Faliszewski2023}, Outcome B reflected the proportions of the underlying population distribution better than Outcome A. 
\end{itemize}

After these explanation treatments (referred to as ``Mechanism'', ``Individual'', and ``Group'' treatment, respectively, in \Cref{sec:explain}), participants were asked how well they understood Method A and Method B. We then prompted them once more to evaluate the methods in terms of fairness and trustworthiness. See \Cref{fig:ex-mechanism}, \Cref{fig:ex-indi}, and \Cref{fig:ex-group} in the Appendix for screenshots of these displays.
\label{sec:design-explain}
\section{Results: Voting Input Formats}
\label{sec:input}

\subsection{Individual Voting Patterns}

\begin{table}[t]
\centering
\caption{Number of projects selected from 24 projects by voters across different voting input formats (for point distribution formats, the number of projects receiving 1 or more points). The table displays the mode (along with its frequency), the mean, median, and standard deviation for each voting input format based on 180 responses.}
\begin{tabular}{l l l r c c}
\toprule
Acronym & Input Format & Mode (Freq) & Mean & Median & Std. Dev. \\ 
\midrule
SN & Select any number of projects & \quad 8 (20) & 10.60 & 10 & 4.73 \\ 
S5 & Select 5 projects & \quad 5 (180) & 5.00 & 5 & 0.00 \\ 
D5 & Distribute 5 points & \quad 4 (73) & 3.51 & 4 & 1.02 \\ 
D10 & Distribute 10 points & \quad 5 (49) & 5.24 & 5 & 1.85 \\ 
S5R & Select 5 and rank & \quad 5 (179) & 4.98 & 5 & 0.30 \\ 
S5D10 & Distribute 10 points for 5 projects & \quad 5 (146) & 4.67 & 5 & 0.81 \\ 
\bottomrule
\end{tabular}
\label{tab:num_proj}
\end{table}

\begin{table}[t]
	\centering
	\small
	\caption{Example of a typical voting pattern in our lab experiment (participant ID: 1PW9NBML). In columns SN and S5, the ticks represent approval. For columns D5, D10, and S5D10, the numbers indicate the points allocated to the respective projects. In the S5R column, the numbers signify the order of ranks, with \#1 being the highest rank. The participant's choices align with the modal number of selections observed in each of the six voting input formats.}
	\begin{tabular}{l c c c c c c}
		\toprule
		Project & SN & S5 & D5 & D10 & S5R & S5D10 \\
		\midrule
		17 Bike Lanes on Seefeldstrasse & $\checkmark$ & $\checkmark$ & 2 pts & 4 pts & \#1 & 4 pts \\
		16 Multicultural Festival at Sechseläutenplatz & $\checkmark$ & $\checkmark$ & 1 pts & 3 pts & \#2 & 3 pts \\
		6 More Night Buses to Oerlikon & $\checkmark$ & $\checkmark$ & 1 pts & 1 pts & \#3 & 1 pts \\
		7 Free Open Badi Space in Wollishofen & $\checkmark$ & $\checkmark$ & 1 pts & 1 pts & \#4 & 1 pts \\
		14 More Trees in Bellevue Sechseläutenplatz & $\checkmark$ & $\checkmark$ & & 1 pts & \#5 & 1 pts \\
		22 Sustainable Cooking Workshop with Kids & $\checkmark$ & & & & & \\
		12 Car Sharing System for Young People & $\checkmark$ & & & & & \\
		5 Safe Bike Paths around Irchel Park & $\checkmark$ & & & & & \\
		\bottomrule
	\end{tabular}
	\label{tab:typ}
\end{table}

We begin by comparing how voters voted in the six voting input formats.
\Cref{tab:num_proj} shows how many projects they voted for, in the sense of selecting a project (SN, S5, S5R) or giving 1 or more points to a project (D5, D10, S5D10). 
The unconstrained SN format results in an average selection of 10.6 projects with a considerable standard deviation of 4.73. For S5, the interface enforced a selection of exactly 5 projects. For S5R, it was possible to select fewer than 5 projects, but all but 1 participant followed the written instruction to select (and rank) exactly 5 projects. For D5, voters on average distributed their points over 3.51 distinct projects, with a mode and median of 4 projects (i.e., a plurality of voters gave 2 points to one project and 1 point each to three projects). The D10 format similarly showed a preference for spreading points, but less evenly, with an average distribution over 5.24 projects with 5 being the mode. For S5D10, a quarter of voters gave 0 points to at least one of the selected 5 projects.

\Cref{tab:typ} offers an example of what a typical vote looks like, by showing the answers of a particular participant on all six formats. This participant selected projects exactly in line with the mode number of selections from \Cref{tab:num_proj} for each input format. For instance, under D10, the participant allocated points primarily to presumably more prioritised projects: ``Bike Lanes on Seefeldstrasse'' received 4 points. In the ranking format S5R, we observe a clear hierarchy in their preferences. Notably, some projects like ``Sustainable Cooking Workshop with Kids'' and ``Car Sharing System for Young People'' were only selected when there were fewer constraints on the selection process. 

Participants generally voted consistently across input formats. For S5, 148 participants (82\%) chose either a superset or a subset of projects they selected for SN. 145 participants (81\%) assigned each project at least as many points under D10 as under D5. 73 participants (41\%) selected the same 5 projects in S5 as in S5R/S5D10, and for 139 participants (77\%) the choice set differed in at most 1 project. For 119 participants (66\%), whenever they ranked a project above another in S5R, they gave it at least as many points in S5D10.

\subsection{Perception of Voting Input Formats}

\begin{figure}[t]
	\centering
	\includegraphics[width=0.8\textwidth]{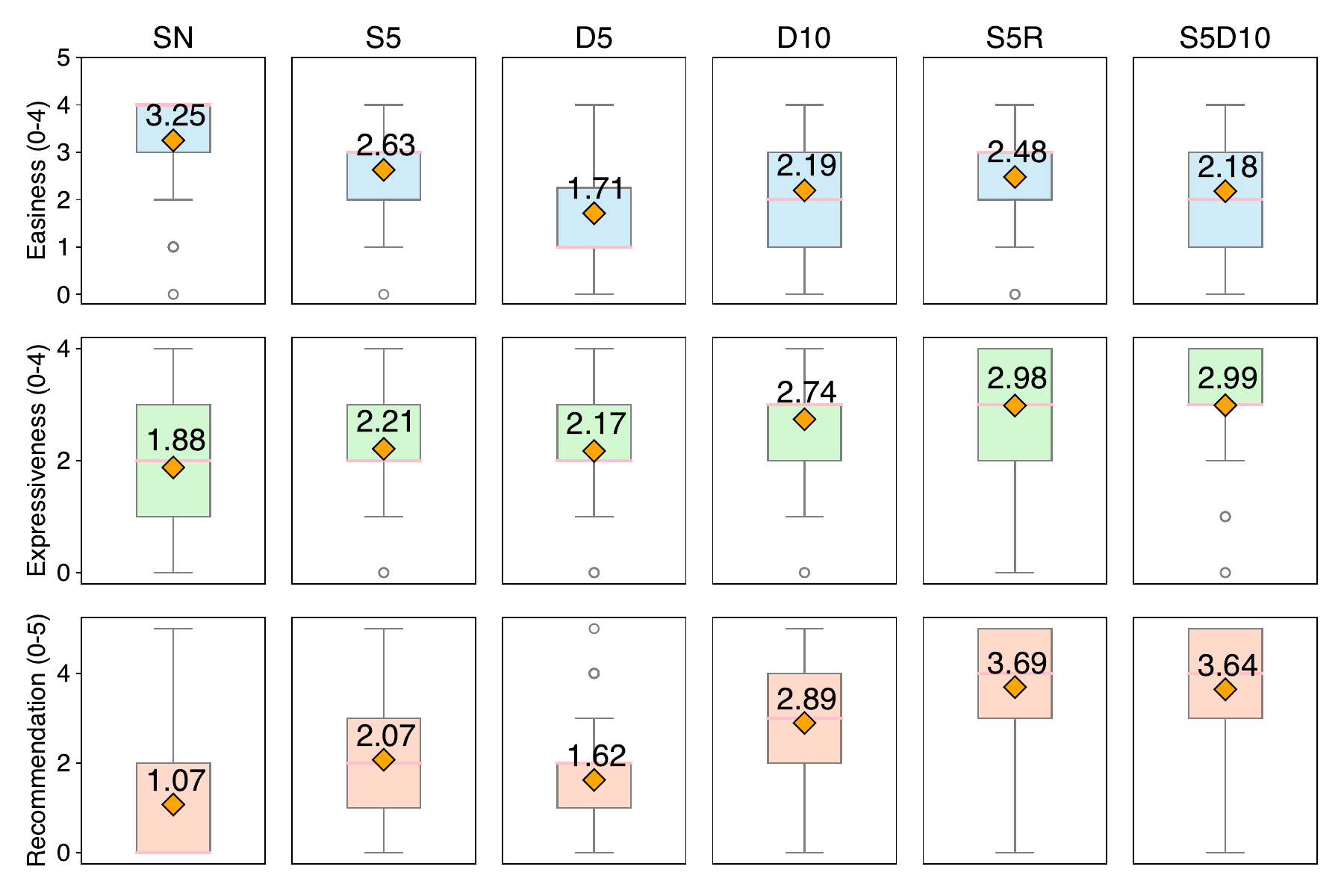}
	\caption{Overview of voter perceptions of different voting formats. The top row shows how voters ranked the formats (Recommendation), quantified via the rank assigned to each format (Borda's Method) with higher values indicating greater recommendation. The middle and bottom rows display the Expressiveness and Easiness, respectively, derived from 5-point Likert scale responses. Box plots indicate the inter-quartile ranges, with medians highlighted in pink and means represented by orange diamonds.}
	\label{fig:pop}
\end{figure}

After the voting process for all six input formats, we asked participants to assess each input format on 5-point Likert scales in terms of Expressiveness and Easiness. The results of a Mann-Whitney U test are presented in \Cref{tab:easiness-rating} and \Cref{tab:expressiveness-rating} of the Appendix.
For Expressiveness, S5D10, S5R, and D10 received high ratings, suggesting that participants felt formats involving a ranking or a distribution of points were representing their preferences better.
For Easiness, SN was viewed as easiest, with a mean rating of 3.25, while D5 was considered the most challenging, with a mean rating of 1.71.

We then asked participants to rank the formats based on their Recommendations for adoption in a city's PB program. For analysis, we translated the rankings into points using Borda's Method. (1st rank: 5 points, 2nd rank: 4 points, and so forth) \parencite{BramsFishburn2002}.
As shown in \Cref{fig:pop}, S5D10 and S5R were ranked highest, obtaining a similar number of points (see also \Cref{table:avg_rank} in the Appendix). Both formats significantly outperformed D10, the third-ranked format (Wilcoxon signed-rank test \(W = 4613\), \(p = 2.8 \times 10^{-7}\)). SN was ranked as least recommended (\(W=5974\), \(p = 0.0017\), see \Cref{table:input-sig} in the Appendix).%
\footnote{To reduce participant workload and ensure a logical progression, all participants used the input formats in the same order. Thus, one might expect that input formats encountered later would be rated higher due to greater familiarity  with the voting process. However, participants actually did not rate later formats as easier to use than earlier formats.}

We performed an OLS regression to see how Expressiveness and Easiness ratings predict the Recommendation rank. (\Cref{tab:reg-rank} in the Appendix). It turns out that Expressiveness is a strong positive predictor of Recommendation with a coefficient of 0.89, while Easiness has a statistically insignificant effect with a coefficient of 0.006. This analysis highlights that Expressiveness, not Easiness of use, appears to be the key factor influencing whether participants recommend an input format.

\subsubsection*{Reasons for Recommendations of Voting Input Formats}
To gain a deeper understanding of the reasons for choosing the ranking of input formats, we analysed the open-ended responses from participants, who were asked to write up their reasons for ranking a format the highest or the lowest. 

We did a thematic evaluation by coding frequent responses and, after this, grouping similar codes into themes. From all the responses obtained, we identified six prevalent themes that explained the reasons for ranking a voting input either first or last.
\Cref{fig:reasons-heat} breaks down how frequently each type of reason was given for each input format.
For the most preferred input format, the main reasons given were clarity (the ability to clearly express preferences), granularity (differentiating degrees of preference), a narrowed choice set (thereby preventing overload), flexibility, ease of use, and clear preference visualisation.  The main reasons given for the least preferred input format were a lack of clarity (inability to prioritise), a lack of granularity (such as too few points for preferences), feeling overwhelmed, lack of specificity, complexity of use, as well as randomness (concern of non-deliberate voting). The detailed explanations of these themes are listed in \Cref{tab:reasons} of the Appendix.

\newcommand{\highreason}[1]{\tikz{\pgfmathsetmacro{\myperc}{#1*1.7}\node [transform shape, rounded corners=1pt,fill=green!65!black!\myperc,inner sep=0, minimum width=18pt, minimum height=8pt] {#1}; }}
\newcommand{\lowreason}[1]{\tikz{\pgfmathsetmacro{\myperc}{#1*0.9}\node [transform shape, rounded corners=1pt,fill=red!75!black!\myperc,inner sep=0, minimum width=18pt, minimum height=8pt] {#1}; }}
\newcommand{\lowreasonwhitetext}[1]{\tikz{\pgfmathsetmacro{\myperc}{#1*0.6}\node [transform shape, rounded corners=1pt,fill=red!80!black!\myperc, text=white, inner sep=0, minimum width=18pt, minimum height=8pt] {#1}; }}

\begin{table}[t]
    \centering
	\caption{Participants used an open-answer text field to explain their rationale for ranking an input format first or last. We categorised these responses into six themes and counted the number of responses for each theme.}
    \begin{subfigure}{0.44\textwidth}
    	\centering
    	\scalebox{0.88}{%
        \begin{tabular}{lcccccc}
        	\toprule
        	& SN & S5 & D5 & D10 & S5R & \clap{S5D10} \\
        	\midrule
        	Clarity & \highreason{4} & \highreason{4} & \highreason{1} & \highreason{11} & \highreason{33} & \highreason{33} \\
        	Granularity & \highreason{1} & \highreason{0} & \highreason{0} & \highreason{14} & \highreason{14} & \highreason{34} \\
        	Narrow Choices & \highreason{0} & \highreason{6} & \highreason{1} & \highreason{1} & \highreason{19} & \highreason{27} \\
        	Flexibility & \highreason{4} & \highreason{1} & \highreason{0} & \highreason{10} & \highreason{4} & \highreason{7} \\
        	Ease & \highreason{3} & \highreason{6} & \highreason{0} & \highreason{3} & \highreason{23} & \highreason{8} \\
        	Visual Repr. & \highreason{1} & \highreason{1} & \highreason{0} & \highreason{0} & \highreason{6} & \highreason{3} \\
        	\bottomrule
       	\end{tabular}}
        \caption{Reasons for ranking an input format \textit{first}.}
        \label{fig:high-heat}
    \end{subfigure}
    \hfill
    \begin{subfigure}{0.52\textwidth}
        \centering
        \scalebox{0.88}{%
        \begin{tabular}{lcccccc}
        	\toprule
        	& SN & S5 & D5 & D10 & S5R & \clap{S5D10} \\
        	\midrule
        	Lack of Clarity & \lowreasonwhitetext{105} & \lowreason{3}  & \lowreason{9}  & \lowreason{1} & \lowreason{0} & \lowreason{1} \\
        	Not Granular   & \lowreason{3}   & \lowreason{6}  & \lowreason{33} & \lowreason{1} & \lowreason{0} & \lowreason{2} \\
        	Overwhelming    & \lowreason{38}  & \lowreason{3}  & \lowreason{3}  & \lowreason{0} & \lowreason{1} & \lowreason{4} \\
        	Lack of Specificity & \lowreason{19}  & \lowreason{2}  & \lowreason{1}  & \lowreason{2} & \lowreason{2} & \lowreason{1} \\
        	Complexity      & \lowreason{2}   & \lowreason{4}  & \lowreason{2}  & \lowreason{3} & \lowreason{2} & \lowreason{1} \\
        	Randomness      & \lowreason{9}   & \lowreason{0}  & \lowreason{0}  & \lowreason{0} & \lowreason{0} & \lowreason{0} \\
        	\bottomrule
       	\end{tabular}}
        \caption{Reasons for ranking an input format \textit{last}.}
        \label{fig:low-heat}
    \end{subfigure}
    \label{fig:reasons-heat}
\end{table}

S5R and S5D10 were ranked highest most often, with participants justifying this due to these formats' clarity and granularity. SN was ranked lowest most often, with participants criticising a perceived lack of clarity.
The results suggest that participants desire the ability to express the granularity of their preferences, but they also seek clear instructions on how to convey these nuances. Too much freedom in expressing preferences can be perceived as overwhelming.

\subsection{Drivers of Voting Behaviour Across Different Voting Input Formats}

\begin{figure}[t]
	\centering
	\includegraphics[width=0.75\textwidth]{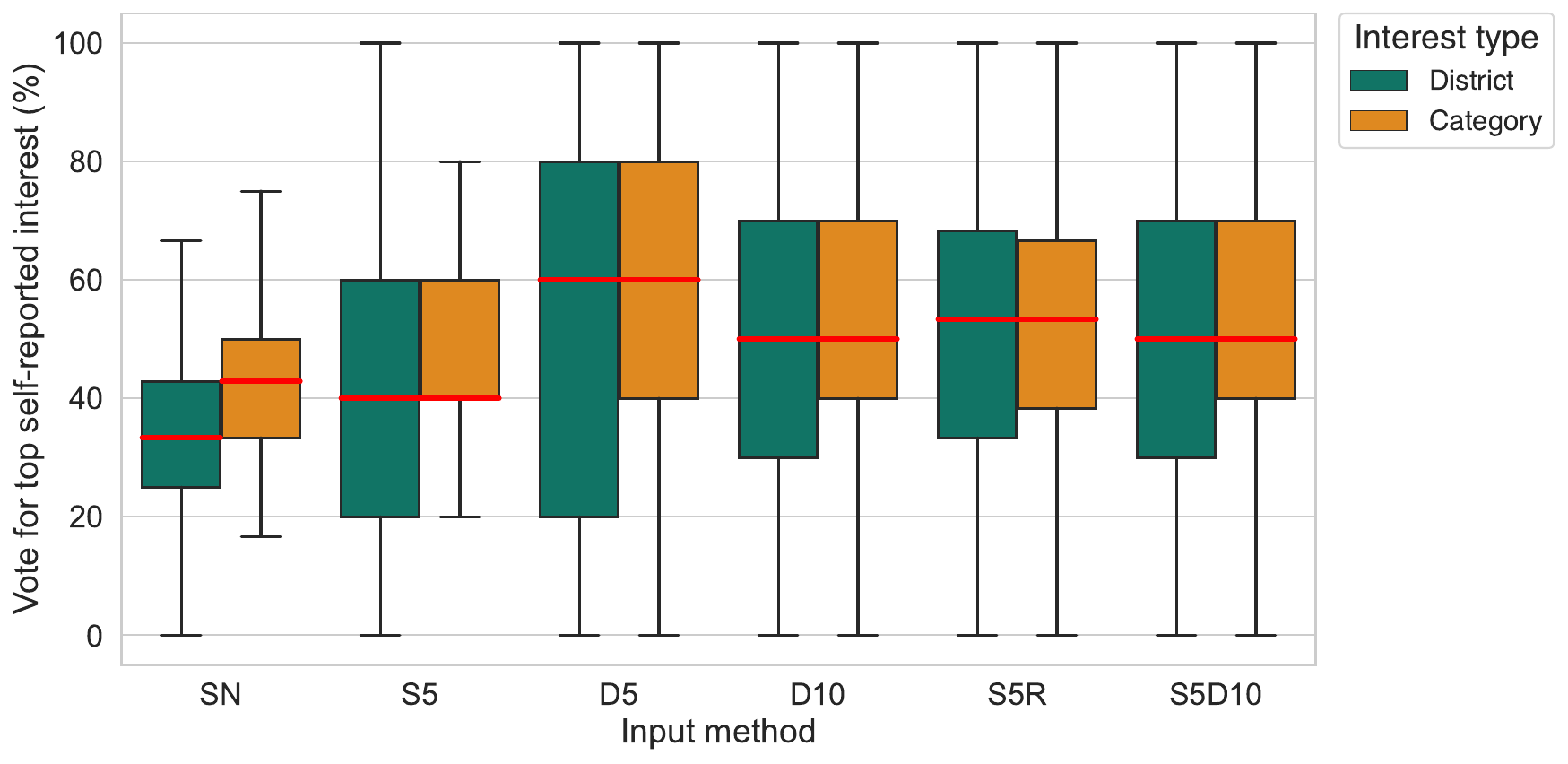}
	\caption{Box plots illustrating the percent of votes that participants assigned to their self-identified preferred district and project category. The medians are highlighted in red.}
	\label{fig:self}
\end{figure}

Before the voting, we asked voters to identify the district and project category they most identified with. To what extent do these preferences drive their votes? \Cref{fig:self} shows what fraction of points participants placed on projects from their identified district and from their preferred category, across various input formats. To compute these fractions, we assume SN and S5 to assign 1 point to each selected project, and S5R to assign points according to Borda's Method. The detailed mean, median and standard deviation statistics are shown in \Cref{tab:self} of the Appendix. For each input format, we note a similar or equal median of the fraction given to district and category projects. Since there were 4 districts and 3 categories, i.e., more projects for each category than for each district, this suggests a stronger emphasis on locations than on topic preferences among voters. Looking at the input formats individually, approval voting formats such as SN and S5 appear to induce less focus on projects with favoured characteristics. In contrast, voting input formats based on point distribution display a higher median voting percentage on their own district or category, all 50\% and above. This concentration of votes is particularly pronounced for D5, with 60\% of the votes allocated to their own districts and category. This may be due to the limitation of having only 5 points in D5, where 33 out of 180 voters also reported to have ``insufficient points to capture preferences accurately'' from the previous thematic analysis (\Cref{fig:low-heat}). In D10, where voters had more points to distribute, the concentration was reduced.

In summary, approval voting tends to encourage voters to disperse their votes beyond their top self-reported interests. Conversely, point-distribution and ranking systems apparently encourage voters to allocate more weight to their top self-reported interests.

We also asked participants to self-report which project characteristics influenced their project selections, using a five-point scale from ``not at all important'' (assigned a value of 1) to ``extremely important'' (value 5). Project category and location were assigned high ratings on average (4.13 and 3.29), while cost and a project's likelihood of winning were seen as much less important (with mean ratings of 2.04 and 1.99).
\subsection{Differences in Voting Outcomes Using Different Voting Input Formats}

\begin{figure}[t]
        \begin{minipage}{0.53\textwidth}
        \centering
        \includegraphics[width=0.95\textwidth]{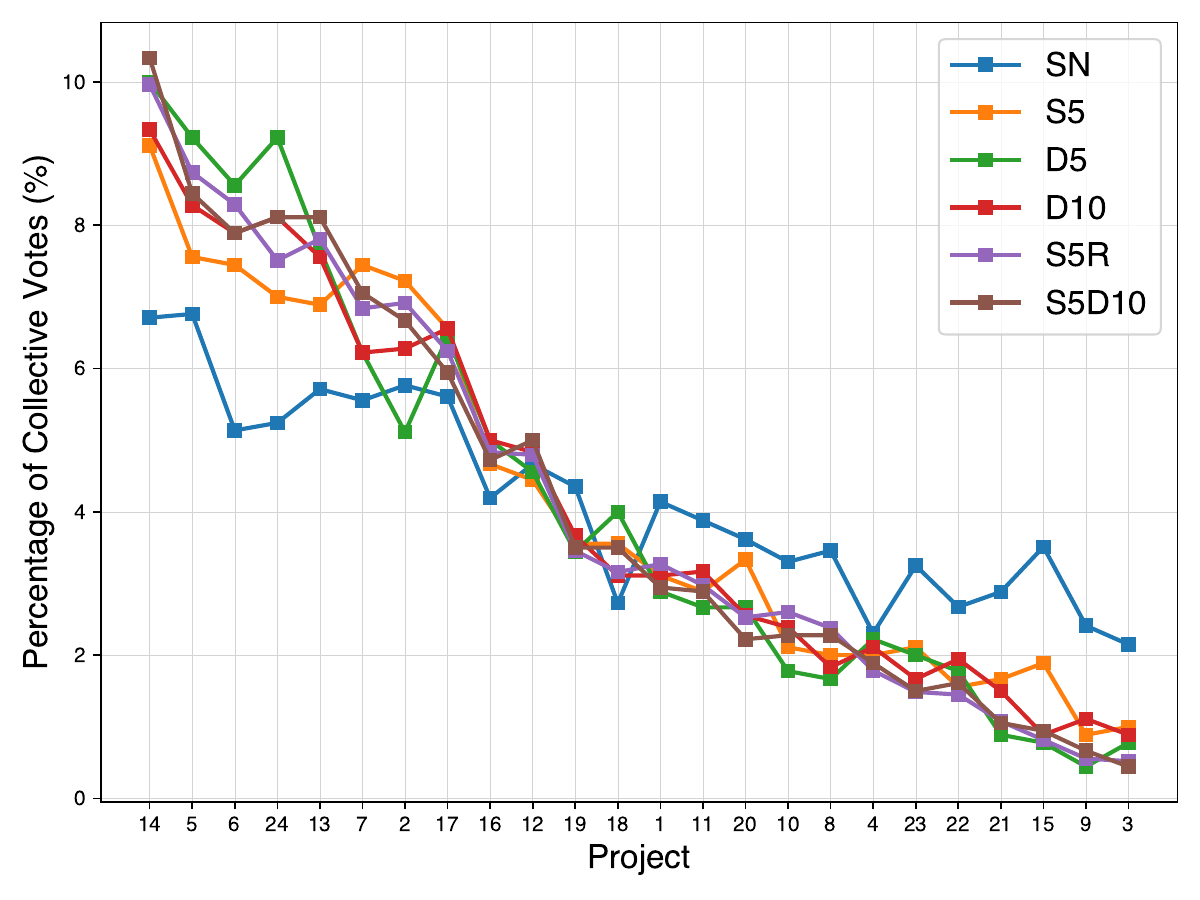}
        \caption{Voting distribution across projects based on six distinct input formats (SN, S5, D5, D10, S5R, S5D10). The projects are ordered by their mean popularity across the 6 inputs.}
        \label{fig:input-dist}
    \end{minipage}%
    \hfill
    \begin{minipage}{0.45\textwidth}
        \centering
        \includegraphics[width=0.95\textwidth]{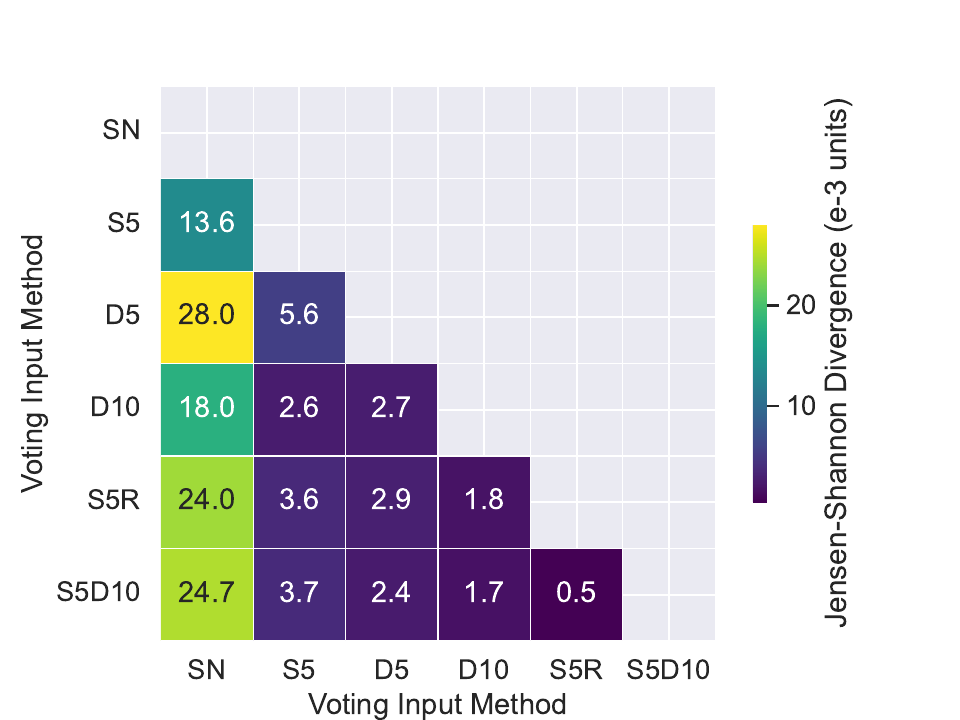}
        \caption{Heat map of Jensen-Shannon divergence values comparing the distribution of votes between different voting input formats. The higher the value, the larger the difference between the distributions.}
        \label{fig:jsdivergence-heatmap}
    \end{minipage}
\end{figure}

Does the choice of the voting input format affect the outcome of the vote? This is an important question to answer, to make an informed decision about input formats. We study this question from two perspectives: first, we consider the summed number of votes or points assigned to the projects and how they differ between formats; second, we compute the outcomes of the Greedy and MES aggregation rules and compare them.

\subsubsection*{Distribution of Votes and Points}

For each of the six voting input formats, we compute the total number of votes or points assigned to each project, by summing across all 180 votes. For the approval formats (SN, S5), projects receive 1 point for each approval. For the point distribution formats (D5, D10, S5D10), projects receive the number of assigned points. For ranking (S5R), we use Borda's Method as described in \Cref{sec:design:inputs}. Then, we normalise the total points of the projects by dividing by the sum, to obtain a point \emph{distribution} indicating the percentage of votes that each project received.

\Cref{fig:input-dist} arranges projects in descending order based on the average percentage of votes they received across input formats, from left to right. We then plot the distribution induced by each of the input formats. Notably, five of the six input formats give a very similar distribution, with SN being the outlier and giving a much flatter distribution (and in particular not as readily identifying the strongest projects in terms of the overall vote share). Visually, the plot suggests that project rankings are consistent across these input formats, implying that the collective preference remains largely unaffected by the format used.

To quantify this difference, as a measure of their (dis)similarity we computed the \href{https://en.wikipedia.org/wiki/Jensen%E2%80%93Shannon_divergence}{Jensen-Shannon (JS) divergence} between distributions (a symmetric version of KL divergence). \Cref{fig:jsdivergence-heatmap} shows the resulting values. SN has a high JS divergence of on average \(18.04 \times 10^{-3}\) to the other input formats, confirming that its distribution of votes is quite distinct from the other formats. On the other hand, the formats ``Distribute 10 points'' (D10) and ``Select 5 projects'' (S5) have low average JS divergences of \(4.46 \times 10^{-3}\) and \(4.85 \times 10^{-3}\) respectively, suggesting they are more similar to other voting formats in terms of the voting distribution. Since S5 is particularly easy to use and to explain to voters, the observation that it induces a similar vote distribution as the more involved formats makes it an attractive choice.

\subsubsection*{Outcomes of Voting Rules}

\begin{figure}[ht]
    \centering
    \begin{minipage}{.48\textwidth}
         \centering
         \includegraphics[width=1.0\textwidth]{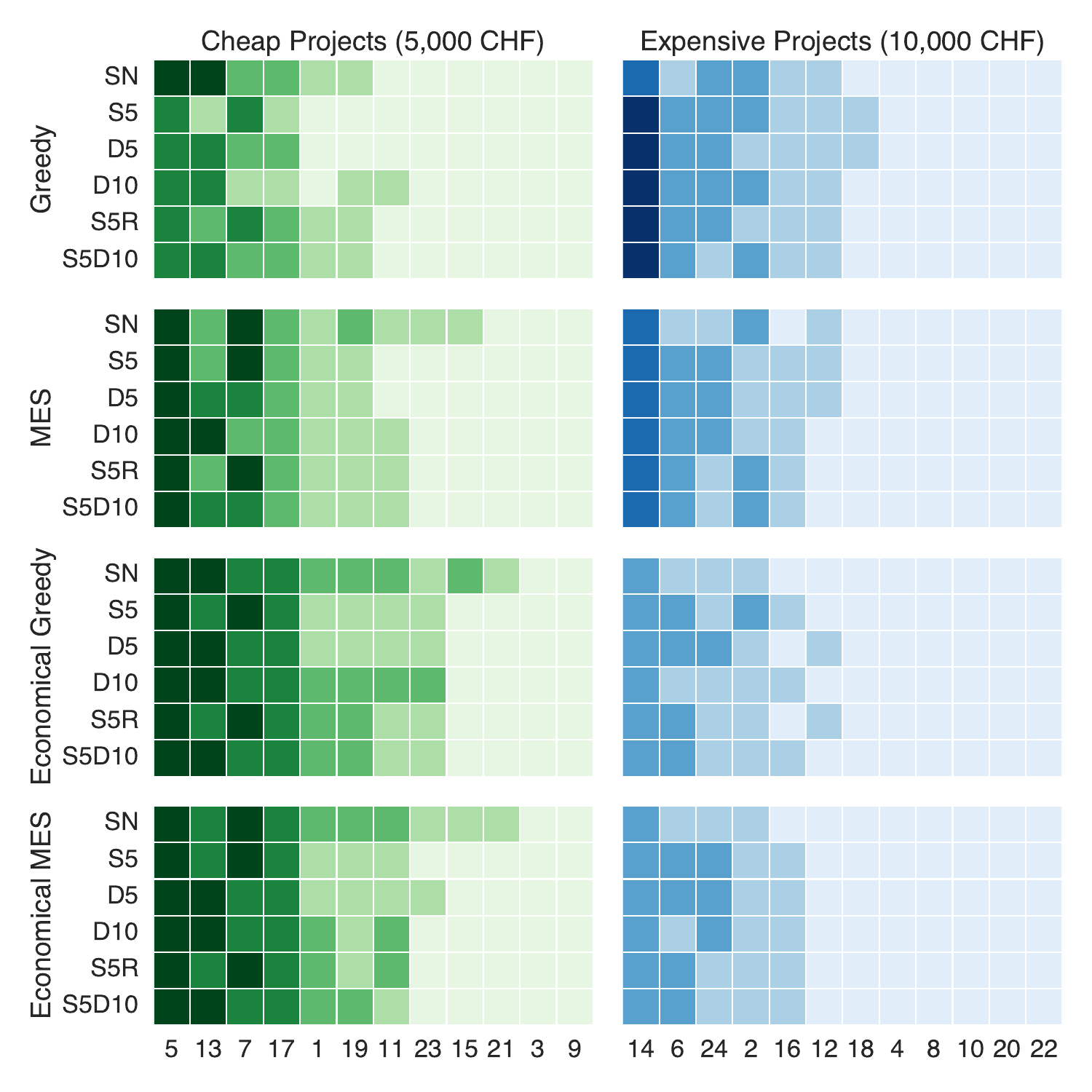}
         \caption{Visualisation of funded projects based on varying voting input formats, aggregation methods, and total budgets. Each row displays which of the 24 projects were selected by a specific combination of input format and aggregation method. The different shades of green or blue represent  the projects that would be funded if the total budget were 10,000, 20,000, 60,000, or 90,000 CHF, from dark to light respectively. Projects with the lightest shade were not funded in all 4 budget scenarios. Projects with a cost of 5,000 CHF are on the left, and those with 10,000 CHF on the right. The heatmap display follows \textcite[Figure 5]{Fairstein2023ParticipatoryWorld}.}
         \label{fig:stability}
    \end{minipage}\hfill
    \begin{minipage}{.48\textwidth}
        \centering
        \includegraphics[width=1.0\textwidth]{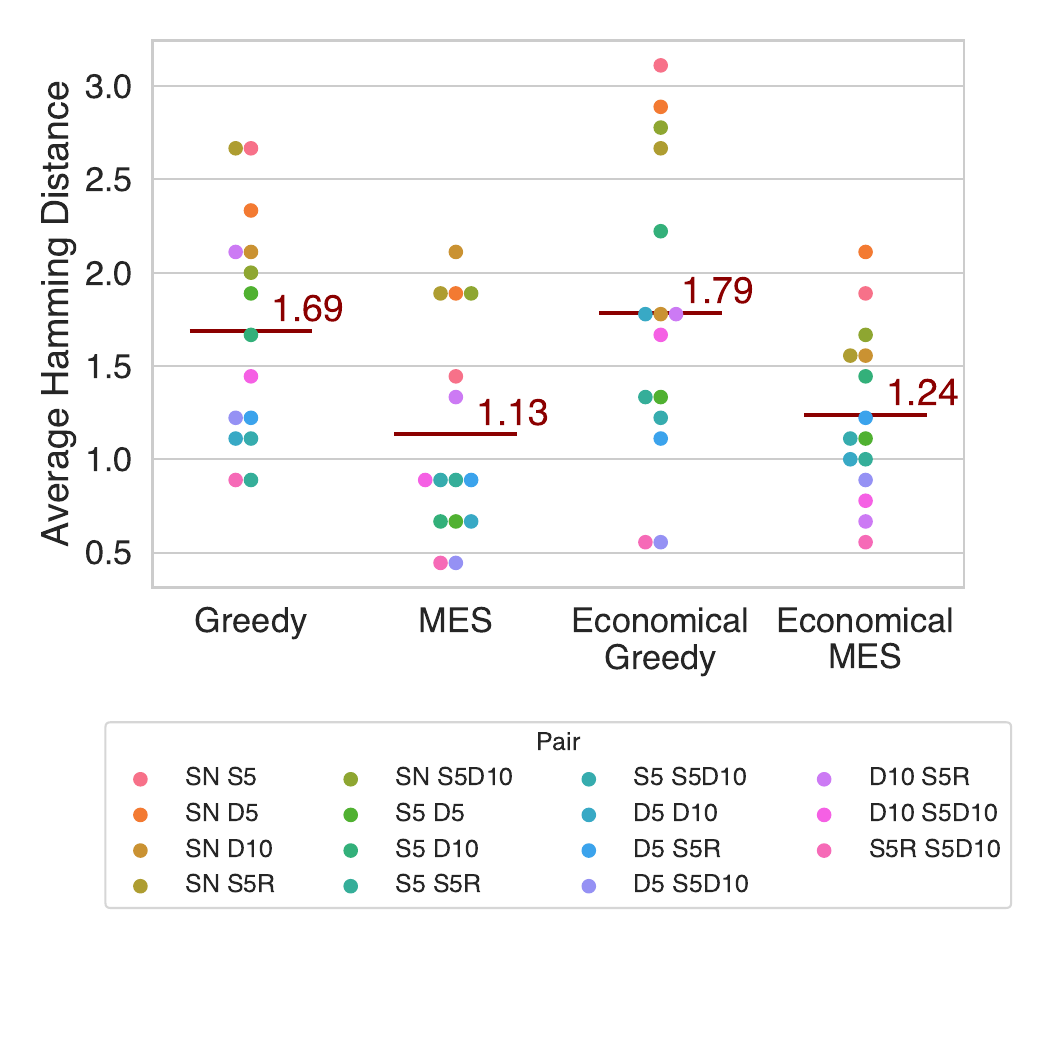}
        \caption{Swarm plot of Hamming distances showing how much the voting outcome according to an aggregation method changes when we change the voting input format. Each point represents the  Hamming distance between a pair of outcomes (as illustrated in \Cref{fig:stability}) obtained from two different voting input formats, averaged over nine values of the budget limit (10,000 to 50,000 CHF in increments of 5,000 CHF). Horizontal lines mark the average Hamming distance for each method, taken over all pairs, thus quantifying the variability under changing input formats for a given aggregation method.}
        \label{fig:hamming}
    \end{minipage}
\end{figure}
We computed the outcomes of the Greedy and MES aggregation methods under various budget settings, as shown in \Cref{fig:stability}. For instance, in the Expensive Projects section, the first column indicates the funding situation of project 14. It is funded by Greedy with a total budget of only 10,000 CHF. In contrast, it is  funded under the MES method only when the budget in the PB program reaches 20,000 CHF. Moreover, if the economical variants of Greedy or MES are applied, project 14 is funded only when the budget reaches 60,000 CHF. We also quantify the difference between the outcome bundles and show it in \Cref{fig:hamming}, with each point indicating the Hamming distance between the outcomes using two different voting inputs.

Examining \Cref{fig:stability,fig:hamming}, we can make the following observations:

\begin{itemize}
	\item \emph{Greedy's Popularity Representation}: As Greedy simply funds the projects with the highest number of votes, its outcome table visually reflects the relative popularity of projects. The roughly equal funding scenario observed between cheaper and more expensive projects suggests that voters were not particularly cost conscious. For example, when the total budget was 90,000 CHF, Greedy with S5 or D5 input would fund 7 expensive projects and only 4 cheap projects in \Cref{fig:stability}. This indicates that voters did not show a stronger preference for cheaper projects when casting their votes and were not overly cost-conscious or frugal in their selections.
	\item \emph{Impact of the Input Format}: Although outcomes that use the SN input format exhibit distinct patterns compared to other input formats, there is no significant evidence suggesting that the choice of input affects the robustness or stability of the outcome in the experiment. Excluding the SN input, variations in input format do not substantially impact which projects are winning. 
    \item \emph{Impact of the Aggregation Methods}: As \Cref{fig:stability} visually demonstrates, Greedy tends to fund more expensive projects at the expense of cheaper ones, whereas the two economical variants are inclined to fund more inexpensive projects, as expected. MES strikes a balance between the two. Regarding the robustness of aggregation methods to varying the voting input format, the average Hamming distance shown in \Cref{fig:hamming} indicates that outcomes using different inputs tend to diverge more when Greedy (1.69) and Economical Greedy (1.79) are employed, compared to MES (1.13) and Economical MES (1.24) which show noticeably lower averages. It is also clear that the choice of the aggregation method influences the outcome more than the choice of the voting input.
\end{itemize}

\section{Results: Aggregation Methods}

In this section, we discuss how participants rated the outcomes produced by the two different aggregation methods: the Greedy rule (``Method A'') and MES (``Method B''). We will analyse how participants rated their fairness and trustworthiness, and how well they understood the methods, as well as how these ratings were influenced by different types of explanations of the outcome.

Translating Likert scales into numerical values from 1 (extremely dissatisfied/unfair) to 5 (extremely satisfied/fair), participants rated their satisfaction with the Greedy outcome to be 2.83 on average, compared to 3.44 for MES, with 79\% of participants rating MES higher than Greedy. For fairness (before explanation), Greedy received a rating of 2.99, and MES received 3.61, with 87\% of participants considering MES fairer than Greedy. After  explanation, fairness ratings were 2.97 for Greedy and 3.93 for MES. Therefore, across both dimensions -- satisfaction and fairness -- participants consistently rated MES higher than Greedy, with all differences being statistically significant (Wilcoxon test, $p<0.001$, see \Cref{tab:aggr_stat} in the Appendix).

\subsection{Correlation Between Utility, Satisfaction, and Perceived Fairness}
\label{sec:corr}
Participants might choose their ratings of voting outcomes according to two conceptually distinct types of reasoning: (1) based on how well they personally liked the projects chosen (e.g., because a voting outcome selected more of the projects they voted for in the first part of the experiment); (2) based on global properties of a voting outcome such as whether it chooses projects with high vote counts or whether it splits the available budget reasonably across districts and categories.

\begin{figure}[t]
    \centering
    \includegraphics[width=\linewidth]{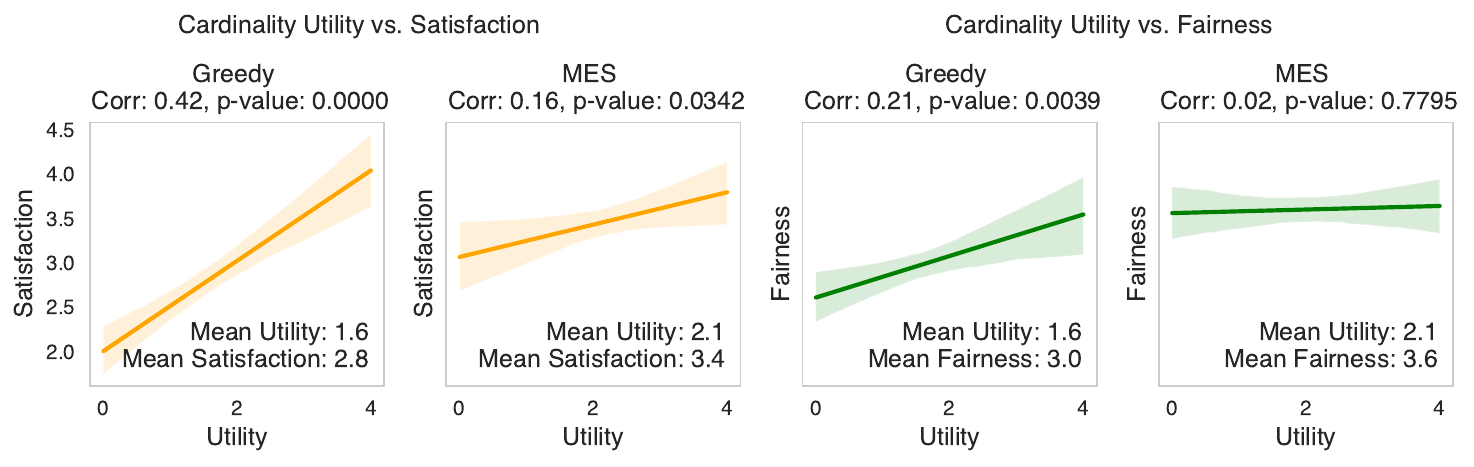}
    \caption{Correlation between Utility and Voter Satisfaction \& Fairness, using S5 voting input and cardinality utility. Mean utility and satisfaction are listed on the bottom right of each plot.}
    \label{fig:utility_corr}
\end{figure}

To distinguish the importance of these two approaches, our analysis in \Cref{fig:utility_corr} shows how their satisfaction and fairness ratings depended on the participants' utility resulting from voting outcomes (measured as the number of projects they selected in the S5 format among those selected for funding). 
This relationship is crucial to understand the legitimacy of participatory budgeting outcomes:
Satisfaction ratings indicate whether the outcomes align with the participants' expectations (outcome legitimacy), while fairness ratings assess the level of equality (throughput legitimacy).

For Satisfaction, \Cref{fig:utility_corr} shows that the Greedy rule outcomes display a stronger correlation between utility and satisfaction compared to MES. For fairness ratings, the outcomes from the Greedy rule suggest that voters tend to perceive an outcome as fairer if it results in higher utility. This significant correlation indicates that, within the Greedy method, fairness perceptions are closely linked to success in terms of utility. In contrast, MES does not show a statistically significant correlation like this, suggesting that perceived fairness associated with MES outcomes is not tied to individual success in terms of utility in the PB program. In the Appendix, we present further results in terms of cost utility, using also formats other than S5, see \Cref{tab:satisfaction_score_correlation,tab:satisfaction_cost_correlation,tab:fairness_score_correlation,tab:fairness_cost_correlation}.

\subsection{Explanations of Voting Aggregation Methods and Outcomes}
\label{sec:explain}

To assess the explainability of the two voting aggregation methods employed, we divided participants into three distinct treatment groups, each receiving different explanations as detailed in \Cref{sec:design-explain}. Afterwards, participants rated how well they understood Method A and Method B, and were asked to rate the two aggregation methods again in terms of fairness and trustworthiness.

Among the ``Mechanism'' participants, it was evident that the underlying algorithm of the Method of Equal Shares was more difficult to grasp, with only 49\% claiming to understand ``a lot'' or ``a great deal'', compared to 87\% for the Greedy method.

Interestingly, despite the relative complexity and lower levels of comprehension of MES within the Mechanism group, its perceived trustworthiness and fairness ratings overall surpass those of the Greedy method. Despite the complexity and lower comprehension levels of MES within the Mechanism group, it garners higher trustworthiness and fairness ratings overall than the Greedy method. 

\begin{figure}[ht]
     \centering
     \includegraphics[width=1\textwidth]{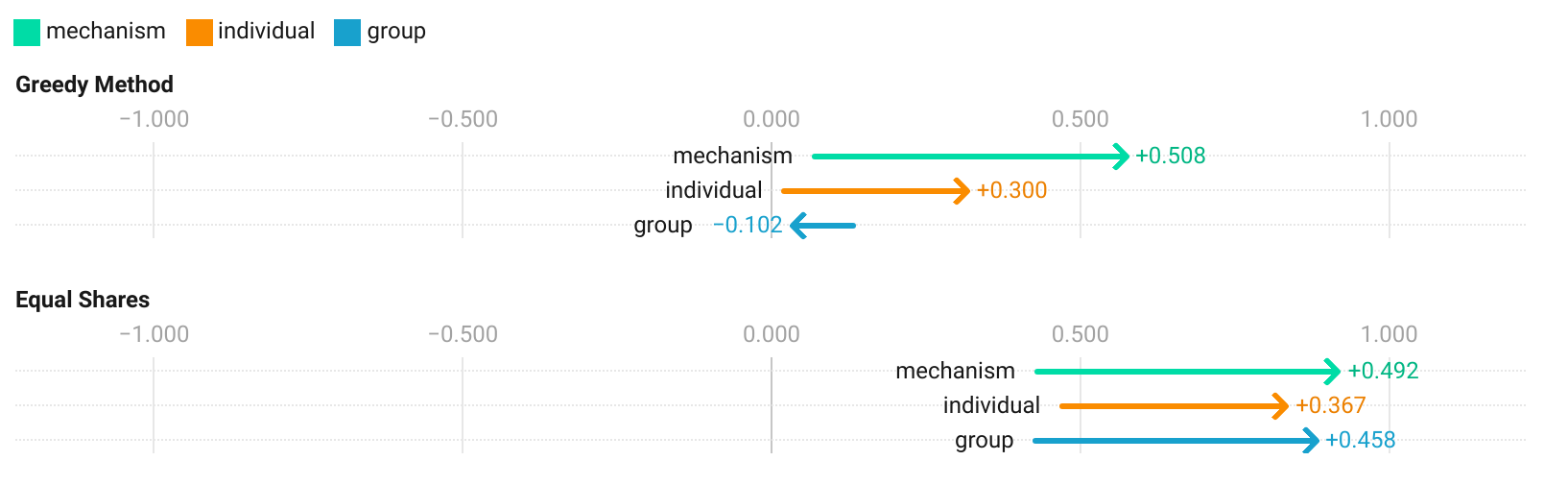}
     \caption{Change in perceived trustworthiness of aggregation methods in different explanation treatment groups.}
     \label{fig:exp-t}
\end{figure}

\begin{figure}[ht]
     \centering
     \includegraphics[width=1\textwidth]{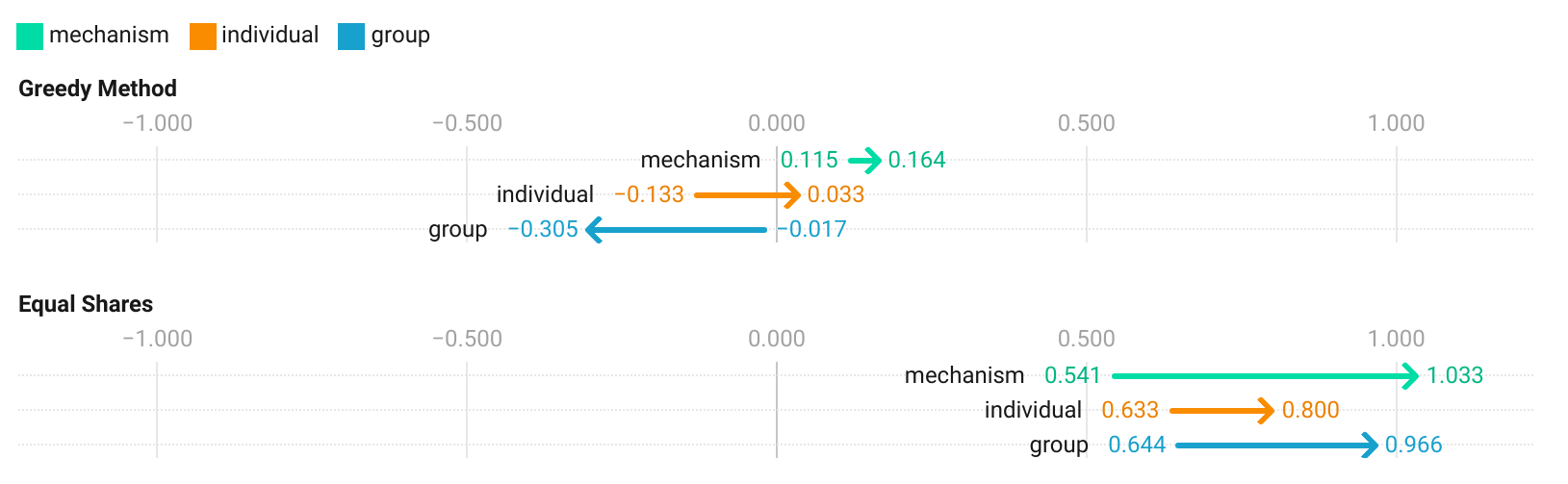}
     \caption{Change in the perceived fairness of aggregation methods in different explanation treatment groups.}
     \label{fig:exp-f}
\end{figure}

The results represented in \Cref{fig:exp-t,fig:exp-f} show the difference in the ratings of fairness and trustworthiness of Greedy and MES before and after the explanation treatment.%
\footnote{Before being shown the explanation, participants rated trustworthiness on a scale ranging from ``strongly disagree'' to ``strongly agree''. After the explanation, the scale ranged from ``not at all trustworthy'' to ``very trustworthy''. Changes in response should be seen in the light of this mismatch. For fairness, the labels were identical before and after the explanation.}
In the figure, we reoriented the 5-point Likert scales to range from -2 to 2, with 0 being a neutral sentiment and a higher rating indicating a more positive evaluation. To have a better understanding of the overall shift of sentiment before and after the explanation, we determined the means of the 180 rating for each question.

\begin{itemize}
	\item In the Mechanism group, to which the detailed workings of the aggregation methods were explained, there was a slight increase in the perceived fairness of the Greedy method post-explanation, from approximately 0.11 to 0.16. However, as shown in \Cref{fig:exp-f}, the perceived fairness of MES improved markedly, from 0.54 to 1.03. Trustworthiness ratings for both methods also show substantial increases, with MES experiencing an increase from 0.43 to 0.92 (\Cref{fig:exp-t}). This is the most effective explanation in terms of boosting the trustworthiness of both algorithms.
	\item In the Individual treatment, where participants were shown the utility distributions among the (simulated) voters, the Greedy method's initially negative fairness rating shifted to a slightly positive rating after the explanation. In contrast, MES saw a small increase from 0.63 to 0.80. However, trustworthiness for both methods showed a substantial improvement.
	\item In the Group treatment, where the budget distribution was shown over districts and categories, the results were different. The Greedy method's fairness rating fell from a near-neutral rating to -0.31 post-explanation, while the fairness rating of MES showed an improvement from 0.64 to 0.97 (\Cref{fig:exp-f}). Trustworthiness for the Greedy method declined slightly, while the trustworthiness of MES improved from 0.42 to 0.88 (\Cref{fig:exp-t}).
\end{itemize}

In summary, while the details provided in the explanations of the different voting methods varied across the three treatment groups, one can observe a consistent trend in \Cref{fig:exp-f,fig:exp-t}: the post-explanation trustworthiness and fairness ratings of MES are higher across all treatment groups.

A further insight from the data is that higher perceived trustworthiness does not necessarily equate to increased perceived fairness. Despite the overall increase in trustworthiness ratings post-explanation, fairness ratings varied. 
This is especially the case in the Mechanism group. While understanding the Greedy method significantly increases the trustworthiness of the method, the fairness rating remains around the same level. This suggests that, increasing ``model transparency'' and a better understanding of a system can contribute to the trustworthiness of an algorithm, but it does not always lead to increased perceptions of fairness. 

\section{Conclusions and Future Work}

\subsection{Discussion}

Our study provides critical insights for the design of digital PB systems, with a particular focus on how different aspects of the voting process affect both the outcomes and the participants' perceptions of these. 

Regarding voting input formats, we found that, in the multi-winner election setting, there are differences from the outcome perspective between, on the one hand, an unstructured voting input format (SN, selecting any number of projects) and, on the other hand, semi-structured voting input formats (S5, D5, D10, S5R, S5D10). However, the choice between different semi-structured voting input formats (ranked or cumulative formats) have a less pronounced effect on the outcome. This supports the findings of \textcite{Fairstein2023ParticipatoryWorld} in their comparison between $k$-approval voting and other voting inputs. However, our findings reveal that voters have strong preferences between voting input formats. Our experimental participants told us that permitting voters to select any number of projects came with a lack of clarity, and that the format of distributing 5 points may lack granularity. Participants showed a preference for voting input formats using rankings or a higher number of points (e.g., 10), suggesting that these formats offer a stronger sense of engagement in the participatory process. The most favoured formats involved ranking selected projects (S5D) or allocating points among selected projects (S5D10). These formats strike a balance between structured decision-making and choice flexibility, making them suitable as recommendations for digital participatory budgeting interfaces that prioritise voter perception in collective decision-making.

Regarding voting aggregation rules, voters showed a strong intuition for outcomes that provide proportional representation and prioritise fairness. These features were significantly correlated with the overall satisfaction with an outcome. Accordingly, there is now experimental support for recent efforts in academia to design, study, and advocate collective decision-making mechanisms that enhance fairness and proportionality, such as MES. When we asked participants to compare the outcomes of the Greedy rule and of MES, they more readily understood the Greedy outcome. However, this difference was reduced after showing participants an explanation or key statistics about the outcomes. Participants tended to assign higher satisfaction ratings to outcomes which funded more of the projects that they personally supported. Furthermore, they tended to assign high fairness ratings to the MES outcome independent of their personal utility, thereby arguably promoting a more community-centric perspective. This observation also suggests that trustworthiness, fairness, and satisfaction are interconnected, hinting at the broader notion that legitimacy is a latent construct that emerges from the combination and interaction of various factors.

In terms of explaining aggregation algorithms, compared to Greedy, MES was perceived as more complex but more trustworthy and fair. Our results also suggest that, in the context of PB, pre-hoc explanations are more effective in improving trust in algorithms than post-hoc explanations. Additionally, increasing ``model transparency'' can contribute to an algorithm's trustworthiness, but does not always lead to enhanced perceptions of fairness, especially when a less fair algorithm becomes more transparent to voters through explanation.

Overall, the voting input and voting aggregation can be seen as the ``front-end'' and ``back-end'' of a voting process, respectively. While the ``front-end'' voting input format may not change the overall result as much, it influences voter perception of their contribution to the outcome as the main point of interaction. In contrast, the ``back end'' voting aggregation method, though less visible to voters, is critical for a fair outcome. An effective voting system consists of a voting input that allows voters to freely express their choices, an aggregation method that supports proportionality, and an explanation that clearly explains both the mechanism and the outcome of the voting process.

\subsection{Practical Relevance} 

In 2023, the city of Aarau, Switzerland, implemented its first PB program (\href{https://www.stadtidee.aarau.ch}{stadtidee.aarau.ch}), for distributing a 50,000 CHF budget among 33 projects. Preliminary results from our laboratory experiment directly informed the setup of the PB process. In particular, the city decided to use a 10-point voting system similar to D10 and to use the Method of Equal Shares to aggregate votes. About 1,800 of approximately 22,000 residents voted. Before the vote, the city website featured an explanation of how the Method of Equal Shares works, inspired by the explanation in our \emph{Mechanism explanation} treatment. After the vote, the \href{https://www.stadtidee.aarau.ch/abstimmung.html/2114}{city website} showed statistics of how the budget was divided between districts, inspired by the explanation in our \emph{Group explanation} treatment.  The key takeaways included: (1) the Method of Equal Shares funded 17 projects, in contrast to just 7 (on average more expensive) projects which would have been chosen by the Greedy method; (2) unlike the Greedy method, the Method of Equal Shares allowed all districts to receive some portion of the budget.

\subsection{Limitations and Future Work}

One potential limitation of our study is the non-representativeness of our participant sample, as all participants were university students in Zurich, which may somewhat limit the generalisability of our findings. Another potential limitation is the lack of full randomisation in the design of the experiment. Given the high complexity of the experiment, in order to reduce the cognitive load and avoid confusion, we chose not to randomise the order of voting input methods and the displayed order of projects. This decision might inherently introduce some bias into participants' perceptions. Furthermore, the simulated nature of our fictional voting may not fully capture the complexities of real-world PB processes. As a result, some of our findings could, in principle, be context-sensitive, affected by specific interface designs, the setting of random outcomes, or the number of projects.

Moving forward, a multifaceted approach to research is recommended:
\begin{itemize}
    \item Continue connecting Social Choice theories to experimental case studies and real-world applications. In particular, we look forward to the publication of the in-depth evaluation of case study in Aarau, Switzerland.
    \item Extend the research to larger and more diverse populations, and to various socio-economic and cultural contexts, thereby increasing the generalisability of our findings.
    \item Investigate the impact of different interface designs and presentation formats on voter engagement and decision-making. Understanding these aspects could inform the development of more intuitive and accessible PB platforms.
    \item Conduct longitudinal studies to grasp the evolution of voter behaviour and satisfaction over time, offering insights into the long-term implications and efficacy of different PB designs.
    \item Investigate human priorities, perceptions of project costs, and utility functions in real-world PB settings.
    \item Explore the criteria used by voters to filter projects before voting and the impact of voting rules on these decisions.
\end{itemize}

In conclusion, our study enhances the understanding of digital participatory budgeting systems, especially in relation to the impact of voting input formats and aggregation methods on fairness and voter perception. It points towards the necessity of further empirical research and of applying insights to real-world PB settings. Future work will focus on refining these systems for greater efficacy and inclusivity, while aiming to move from theoretical explorations to practical applications in digital governance.

\subsection{Acknowledgements}
This study was financed by the Swiss National Science Foundation (SNSF) as part of the National Research Programme NRP 77 Digital Transformation, project no. 187249 [Applicants: Regula Hänggli Fricker, Evangelos Pournaras, Dirk Helbing]. 
Evangelos Pournaras is also supported by the UKRI Future Leaders Fellowship (MR\-/W009560\-/1): `\emph{Digitally Assisted Collective Governance of Smart City Commons--ARTIO}'. We would further like to thank the Decision Lab of ETH Zürich for their great support in carrying out the lab experiment. Special thanks also go to the city of Aarau, especially the Aarau Stadtidee team, and Thomas Wellings from the team from the University of Leeds for their insightful suggestions and their commitment to scientific collaboration. Last but not least, we would like to thank three anonymous reviewers for their useful feedback that helped us improve the presentation of our paper.

\newpage
\printbibliography
\newpage
\appendix

\section{Additional Tables}

\subsection{Voting Input Formats}

\begin{minipage}[b]{0.45\linewidth}
	\begin{table}[H]
		\centering
		\caption{Mann-Whitney U test results for easiness rating on input formats.}
		\small
		\label{tab:easiness-rating}
		\begin{tabular}{l l l}
			\toprule
			Comparison & $U$-statistic & $p$-value \\
			\midrule
			SN vs S5 & 22696.0 & $3.71\times 10^{-12}$ \\
			SN vs D5 & 26650.0 & $5.54\times 10^{-28}$ \\
			SN vs D10 & 24857.5 & $8.25\times 10^{-20}$ \\
			SN vs S5R & 22880.5 & $1.25\times 10^{-12}$ \\
			SN vs S5D10 & 24518.5 & $1.86\times 10^{-18}$ \\
			S5 vs D5 & 23512.5 & $1.37\times 10^{-14}$ \\
			S5 vs D10 & 20198.0 & $2.09\times 10^{-05}$ \\
			S5 vs S5R & 17424.0 & 0.19200 \\
			S5 vs S5D10 & 20069.5 & $4.29\times 10^{-05}$ \\
			D5 vs D10 & 12056.0 & $1.29\times 10^{-05}$ \\
			D5 vs S5R & 10204.5 & $3.61\times 10^{-10}$ \\
			D5 vs S5D10 & 12376.0 & $6.01\times 10^{-05}$ \\
			D10 vs S5R & 13717.0 & 0.00907 \\
			D10 vs S5D10 & 16341.5 & 0.88216 \\
			S5R vs S5D10 & 18670.0 & 0.00974 \\
			\bottomrule
		\end{tabular}
	\end{table}
\end{minipage}
\hfill 
\begin{minipage}[b]{0.45\linewidth}
	\begin{table}[H]
		\centering
		\caption{Mann-Whitney U test results for the rating of Expressiveness as a function of  input formats.}
		\small
		\label{tab:expressiveness-rating}
		\begin{tabular}{l l l}
			\toprule
			Comparison & $U$-statistic & $p$-value \\
			\midrule
			SN vs S5 & 13260.0 & 0.00213 \\
			SN vs D5 & 13515.5 & 0.00501 \\
			SN vs D10 & 9998.5 & $8.99\times 10^{-11}$ \\
			SN vs S5R & 8757.0 & $9.68\times 10^{-15}$ \\
			SN vs S5D10 & 8548.0 & $1.58\times 10^{-15}$ \\
			S5 vs D5 & 16823.0 & 0.50425 \\
			S5 vs D10 & 10849.0 & $7.99\times 10^{-09}$ \\
			S5 vs S5R & 8903.0 & $1.37\times 10^{-14}$ \\
			S5 vs S5D10 & 8761.0 & $2.41\times 10^{-15}$ \\
			D5 vs D10 & 10444.0 & $7.98\times 10^{-10}$ \\
			D5 vs S5R & 8779.5 & $6.25\times 10^{-15}$ \\
			D5 vs S5D10 & 8533.0 & $5.12\times 10^{-16}$ \\
			D10 vs S5R & 13172.5 & 0.00111 \\
			D10 vs S5D10 & 13563.0 & 0.00406 \\
			S5R vs S5D10 & 16737.5 & 0.56367 \\
			\bottomrule
		\end{tabular}
	\end{table}
\end{minipage}

\begin{table}[H]
\centering
\caption{Average ranks and standard deviations of the six voting input formats studied. Smaller average rank values indicate higher popularity. Collected from 180 participants.}
\small
\begin{tabular}{l c c}
\toprule
Input Format & Average Rank & Std. Dev. \\ 
\midrule
Select 5 and rank & 2.31 & 1.32 \\
Select 5 and distribute 10 points & 2.36 & 1.34 \\
Distribute 10 points & 3.11 & 1.39 \\
Select 5 projects & 3.93 & 1.36 \\
Distribute 5 points & 4.38 & 1.20 \\
Select any number of projects & 4.93 & 1.71 \\
\bottomrule
\end{tabular}
\label{table:avg_rank}
\end{table}

\begin{table}[H]
\centering
\caption{Input formats ranked by participants in the order of recommendation for use of PB by a city. The formats were ordered by average rank, and adjacent formats were compared using the Wilcoxon signed-rank test for significant rank differences.} 
\small
\begin{tabular}{l l l}
\toprule
Comparison & $W$-statistic & $p$-value \\ 
\midrule
S5R $\succ$ S5D10 & 7965.5 & 0.7920 \\
S5D10 $\succ$ D10 & 4613.0 & $2.78 \times 10^{-7}$ \\
D10 $\succ$ S5 & 5067.0 & $8.95 \times 10^{-6}$ \\
S5 $\succ$ D5 & 6170.5 & 0.0040 \\
D5 $\succ$ SN & 5973.5 & 0.0017 \\
\bottomrule
\end{tabular}
\label{table:input-sig}
\end{table}

\begin{table}[H]
\centering
\caption{Regression results for Recommendation of voting input formats.}
\small
\begin{tabular}{lcc}
\toprule
Variable & Coefficient & $p$-value \\
\midrule
Constant & 0.266 & 0.043* \\
Expressiveness & 0.890 & <0.001*** \\
Easiness & 0.006 & 0.879 \\
\midrule
\multicolumn{3}{c}{Model Statistics} \\
\midrule
R-squared & \multicolumn{2}{c}{0.327} \\
F-statistic (Prob.) & \multicolumn{2}{c}{261.9 (<0.001***)} \\
\bottomrule
\end{tabular}
\label{tab:reg-rank}
\end{table}

\begin{table}[H]
    \centering
     \caption{Reasons for ranking a voting method as either first or last. Participants provided their rationale using an open-answer text field. These responses were subsequently categorised by the authors into specific themes, and the percentage of responses calculated for each theme.}
     \small
     \label{tab:reasons}
    \begin{tabular}{ll}
        \toprule
        \textbf{First} & \textbf{\%} \\
        \midrule
        Preference Clarity & 48 \\
        Granularity & 35 \\
        Limited choices & 30 \\
        Ease of process & 24 \\
        Flexibility & 14 \\
        Visual Representation & 6 \\
        \midrule
        \textbf{Last} & \textbf{\%} \\
        \midrule
        Lack of Preference Clarity & 66 \\
        Overwhelming Choices & 27 \\
        Limited Allocation & 25 \\
        Lack of Specificity & 15 \\
        Complexity & 8 \\
        Randomness & 5 \\
        \bottomrule
    \end{tabular}
\end{table}

\begin{table}[H]
	\centering
	\caption{Statistical summary of percent of vote going to projects of the participant's selected district or selected preferred category, by input format.}
	\small
	\label{tab:self}
	\begin{tabular}{l c c c c c c}
		\toprule
		& \multicolumn{3}{c}{District} & \multicolumn{3}{c}{Category} \\
		\cmidrule(lr){2-4} \cmidrule(lr){5-7}
		Input & Mean & Median & Std. Dev. & Mean & Median & Std. Dev. \\ 
		\midrule
		SN & 37.42 & 33.33 & 19.61 & 44.76 & 42.86 & 15.04 \\ 
		S5 & 44.78 & 40.00 & 23.58 & 48.33 & 40.00 & 20.91 \\ 
		D5 & 53.22 & 60.00 & 33.50 & 53.33 & 60.00 & 28.36 \\ 
		D10 & 50.39 & 50.00 & 30.02 & 50.61 & 50.00 & 23.09 \\ 
		S5R & 51.33 & 53.33 & 27.15 & 51.22 & 53.33 & 22.68 \\ 
		S5D10 & 50.94 & 50.00 & 29.06 & 51.72 & 50.00 & 24.40 \\ 
		\bottomrule
	\end{tabular}
\end{table}

\subsection{Aggregation Methods}

\begin{table}[H]
	\centering
	\caption{Statistical summary of Satisfaction and Fairness ratings and Wilcoxon test results for MES being rated higher than Greedy.}
	\label{tab:aggr_stat}
	\begin{tabular}{lcccc}
		\toprule
		Metric & Mean & Median & Std. Dev. & $p$-value \\
		\midrule
		Satisfaction (Greedy) & 2.83 & 3.00 & 1.15 & \\
		Satisfaction (MES) & 3.44 & 4.00 & 1.04 & $3.18 \times 10^{-7}$ \\
		\midrule
		Fairness Before (Greedy) & 2.99 & 3.00 & 1.03 & \\
		Fairness Before (MES) & 3.61 & 4.00 & 0.87 & $6.58 \times 10^{-7}$ \\
		\midrule
		Fairness After (Greedy) & 2.97 & 3.00 & 1.07 & \\
		Fairness After (MES) & 3.93 & 4.00 & 0.77 & $2.66 \times 10^{-14}$ \\
		\bottomrule
	\end{tabular}
\end{table}

\begin{table}[H]
\centering
\caption{Correlation (and $p$-value) between satisfaction and cardinality utility for different voting methods.}
\small
\begin{tabular}{lcccccc}
\toprule
Aggregation Method & SN & S5 & D5 & D10 & S5R & S5D10 \\
\midrule
MES & -0.07 (0.3454) & 0.16 (0.0342) & 0.32 (0.0000) & 0.34 (0.0000) & 0.28 (0.0002) & 0.29 (0.0001) \\
Greedy & 0.23 (0.0019) & 0.42 (0.0000) & 0.42 (0.0000) & 0.47 (0.0000) & 0.47 (0.0000) & 0.42 (0.0000) \\
\bottomrule
\end{tabular}
\label{tab:satisfaction_score_correlation}
\end{table}

\begin{table}[H]
\centering
\caption{Correlation (and $p$-value) between satisfaction and cost utility for different voting methods.}
\small
\begin{tabular}{lcccccc}
\toprule
Aggregation Method & SN & S5 & D5 & D10 & S5R & S5D10 \\
\midrule
MES & -0.04 (0.5604) & 0.20 (0.0063) & 0.31 (0.0000) & 0.33 (0.0000) & 0.27 (0.0002) & 0.29 (0.0001) \\
Greedy & 0.26 (0.0003) & 0.43 (0.0000) & 0.43 (0.0000) & 0.49 (0.0000) & 0.48 (0.0000) & 0.43 (0.0000) \\
\bottomrule
\end{tabular}
\label{tab:satisfaction_cost_correlation}
\end{table}

\begin{table}[H]
\centering
\caption{Correlation (and $p$-value) between fairness and cardinality utility for different voting methods.}
\small
\begin{tabular}{lcccccc}
\toprule
Aggregation Method & SN & S5 & D5 & D10 & S5R & S5D10 \\
\midrule
MES & 0.05 (0.4948) & 0.02 (0.7795) & 0.07 (0.3764) & 0.10 (0.1973) & 0.07 (0.3175) & 0.07 (0.3707) \\
Greedy & 0.17 (0.0195) & 0.21 (0.0039) & 0.23 (0.0020) & 0.25 (0.0009) & 0.22 (0.0030) & 0.18 (0.0130) \\
\bottomrule
\end{tabular}
\label{tab:fairness_score_correlation}
\end{table}

\begin{table}[H]
\centering
\caption{Correlation (and $p$-value) between fairness and cost utility for different voting methods.}
\small
\begin{tabular}{lcccccc}
\toprule
Aggregation Method & SN & S5 & D5 & D10 & S5R & S5D10 \\
\midrule
MES & 0.05 (0.5021) & 0.02 (0.8209) & 0.04 (0.6189) & 0.07 (0.3546) & 0.05 (0.5084) & 0.06 (0.4496) \\
Greedy & 0.17 (0.0206) & 0.19 (0.0104) & 0.23 (0.0022) & 0.23 (0.0022) & 0.22 (0.0030) & 0.18 (0.0133) \\
\bottomrule
\end{tabular}
\label{tab:fairness_cost_correlation}
\end{table}

\section{Experiment Interface}

\subsection{Voting Interface}
   
\begin{figure}[H]
   	\centering
    \begin{subfigure}{0.45\linewidth}
        \includegraphics[width=\linewidth]{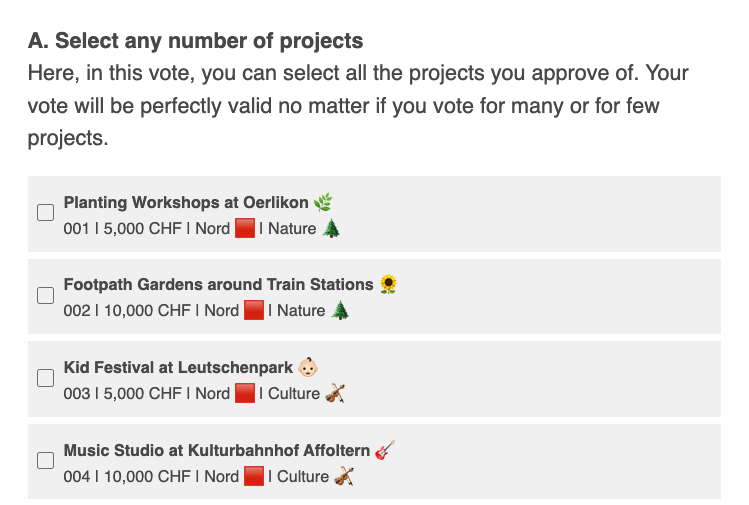}
        \caption{Screenshot of the SN format interface.}
        \label{fig:sn}
    \end{subfigure}
    \hfill
    \begin{subfigure}{0.45\linewidth}
        \includegraphics[width=\linewidth]{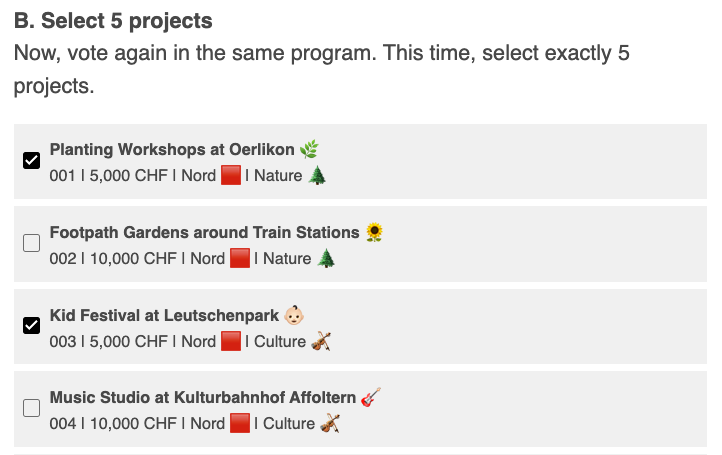}
        \caption{Screenshot of the S5 format interface.}
        \label{fig:s5}
    \end{subfigure}

    \begin{subfigure}{0.45\linewidth}
        \includegraphics[width=\linewidth]{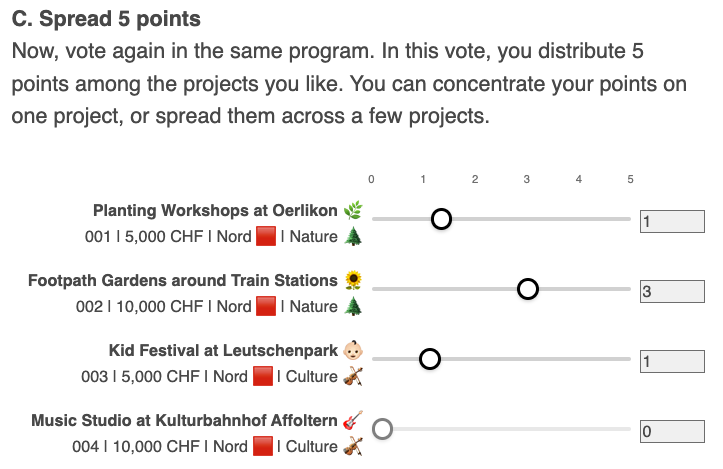}
        \caption{Screenshot of the D5 format interface.}
        \label{fig:d5}
    \end{subfigure}
    \hfill
    \begin{subfigure}{0.45\linewidth}
        \includegraphics[width=\linewidth]{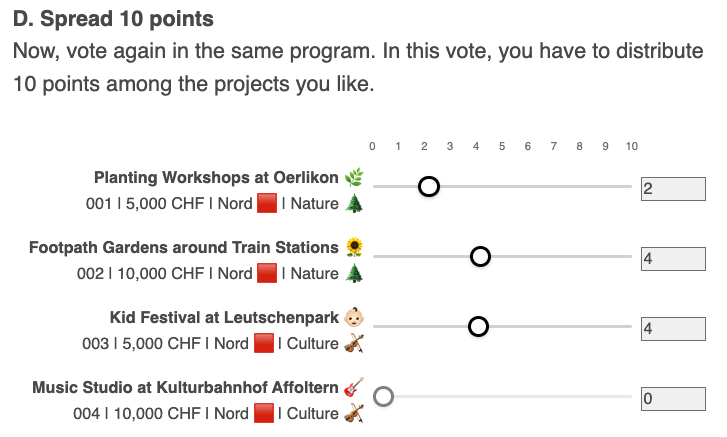}
        \caption{Screenshot of the D10 format interface.}
        \label{fig:d10}
    \end{subfigure}

    \begin{subfigure}{0.45\linewidth}
        \includegraphics[width=\linewidth]{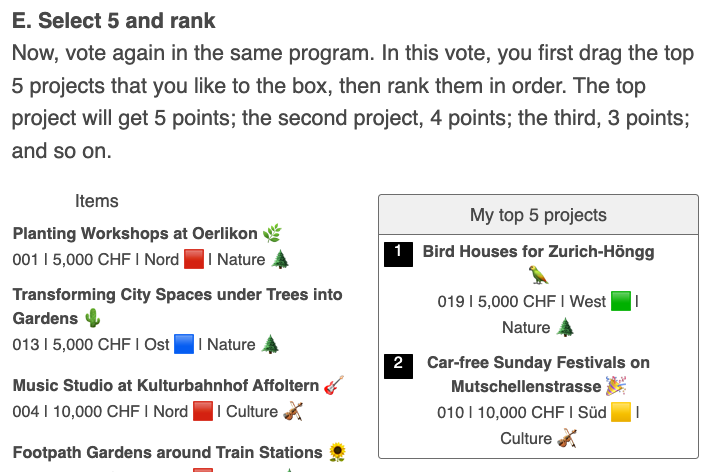}
        \caption{Screenshot of the S5R format interface.}
        \label{fig:s5r}
    \end{subfigure}
    \hfill
    \begin{subfigure}{0.45\linewidth}
        \includegraphics[width=\linewidth]{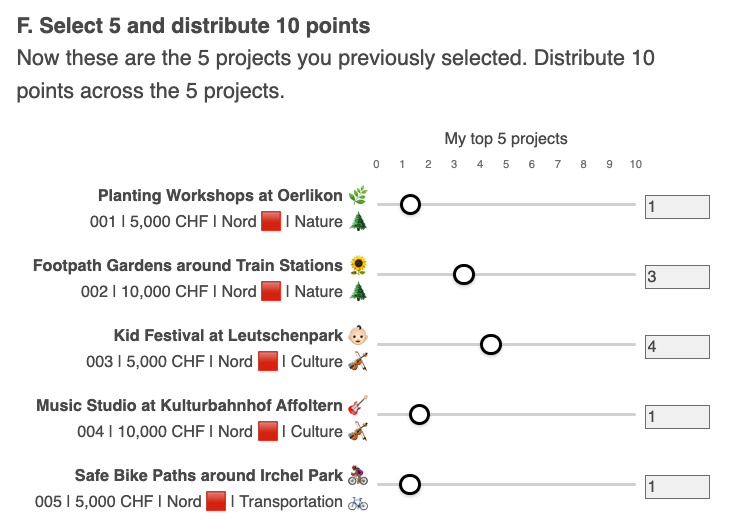}
        \caption{Screenshot of the S5D10 format interface.}
        \label{fig:s5d10}
    \end{subfigure}
    
    \caption{Screenshots of voting input format screens.}
    \label{fig:input-screenshots}
\end{figure}

\subsection{Simulated Voting Outcomes}
\begin{figure}[H]
	\centering
	\includegraphics[width=0.6\linewidth]{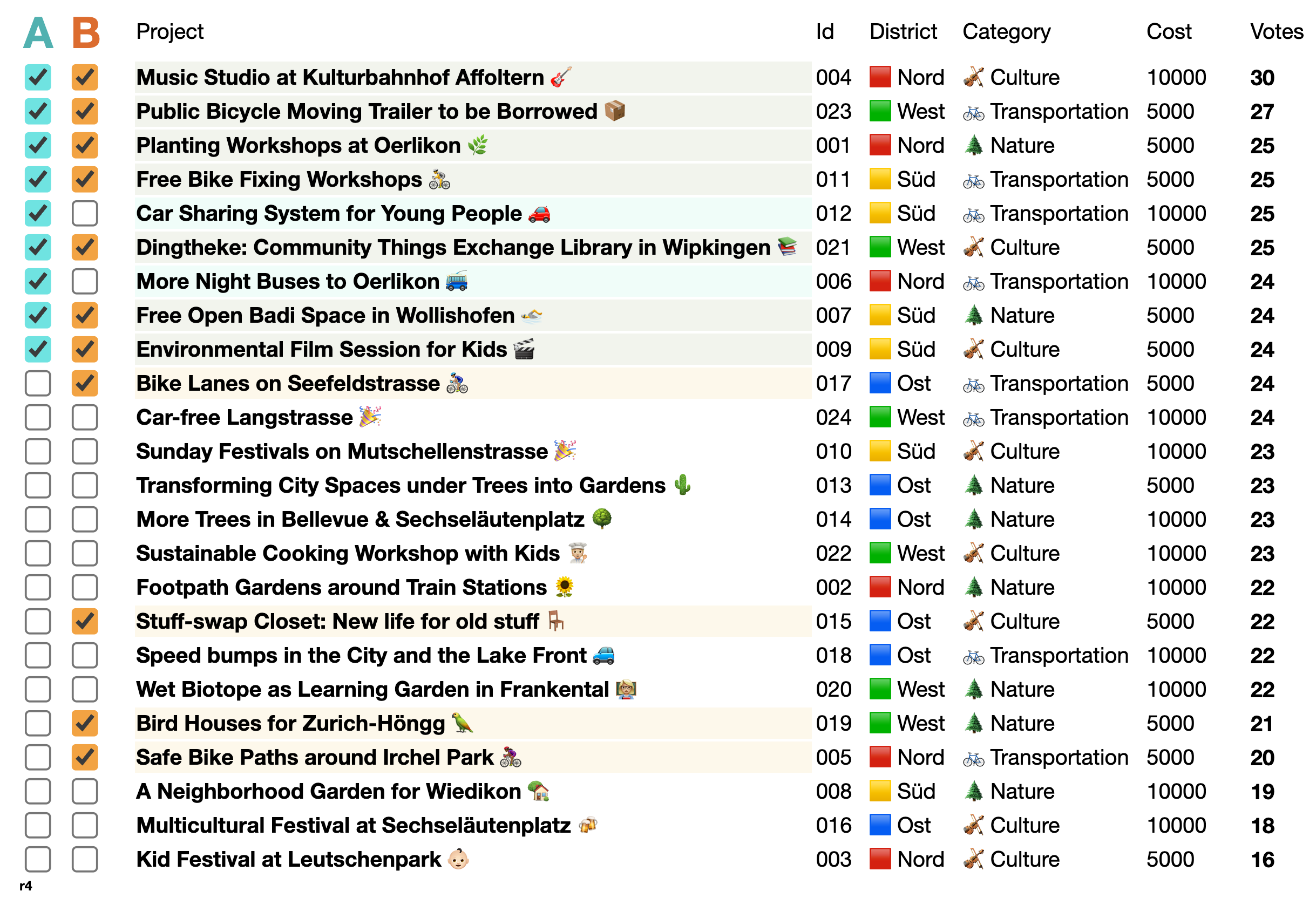}
	\caption{Simulated outcome table for the fictional Zürich PB with voters having randomly sampled preferences (uniform distribution, ``Instance 1''). Columns A and B show the projects selected using the Greedy rule and the Method of Equal Shares (MES), respectively.}
	\label{fig:random-outcome-table}
\end{figure}

\begin{figure}[ht]
	\centering
	\includegraphics[width=0.6\linewidth]{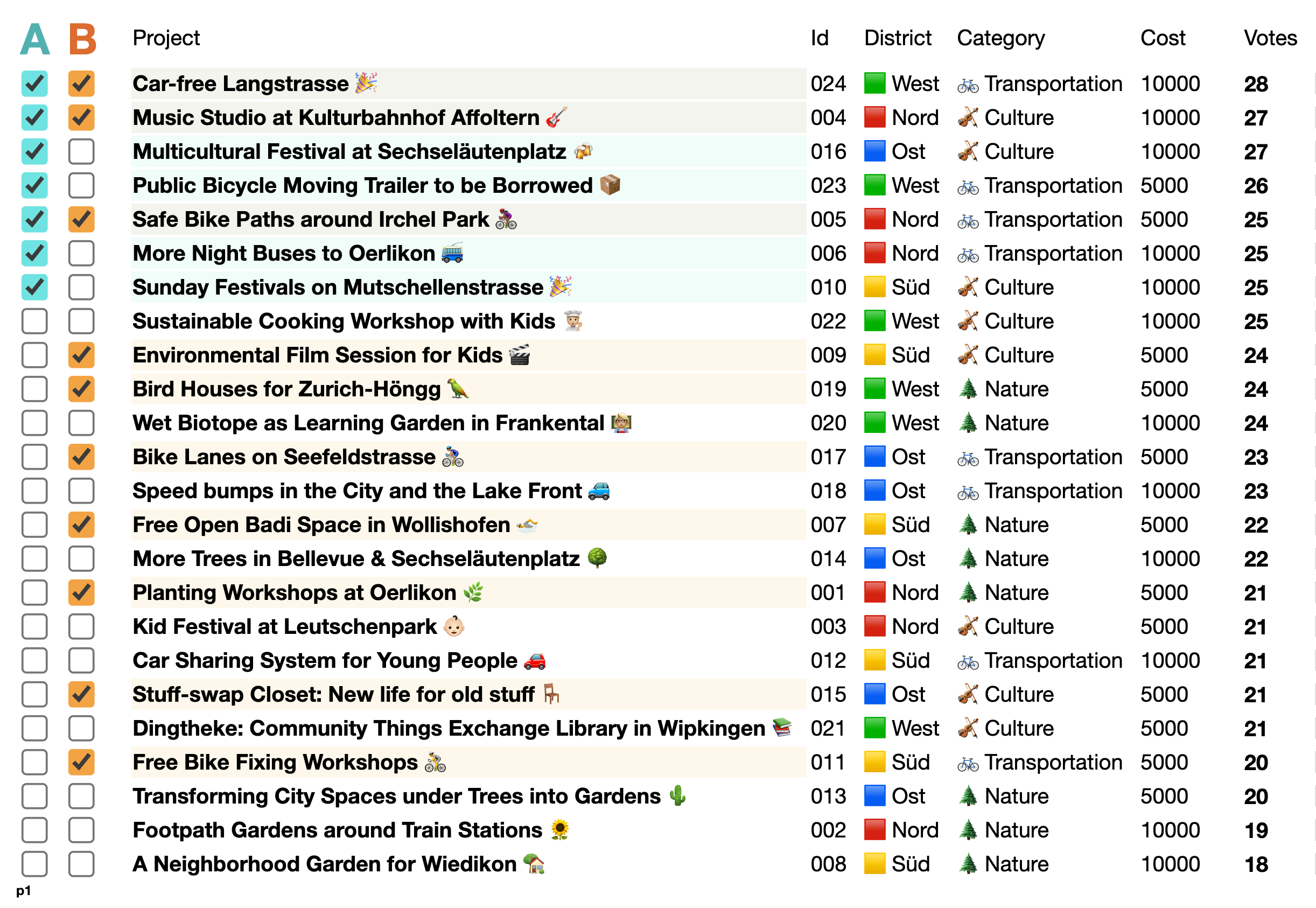}
	\caption{Simulated outcome table for the fictional Zürich PB with voters having randomly sampled preferences (with strong preferences based on district and categories, ``Instance 2''). Columns A and B show the projects selected using the Greedy rule and the Method of Equal Shares (MES), respectively.}
	\label{fig:polarised-outcome-table}
\end{figure}

\subsection{Explanations}

\begin{figure}[ht]
	\centering
	\begin{subfigure}{0.48\textwidth}
		\centering
		\includegraphics[width=\linewidth]{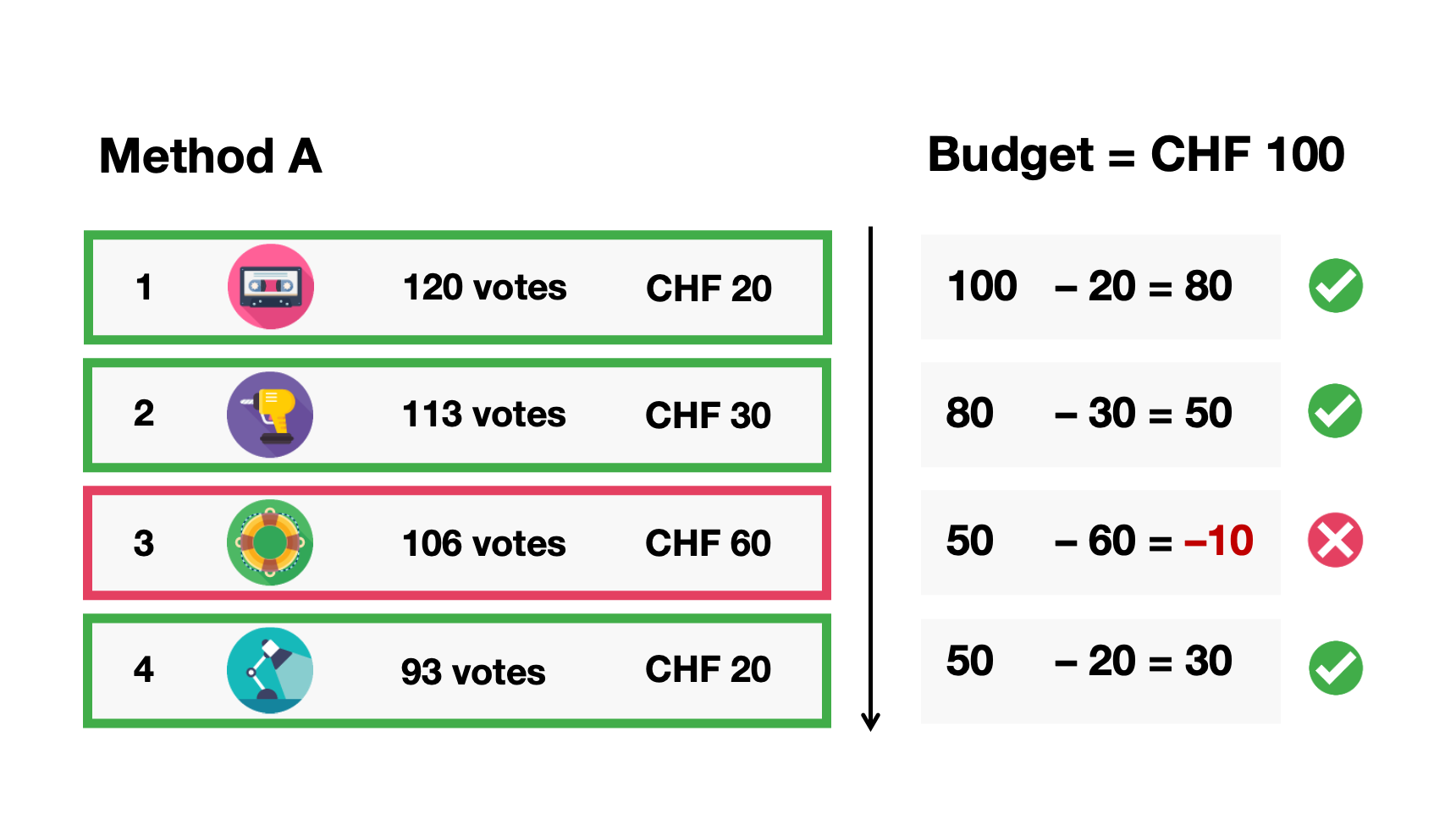}
		\caption{Illustration used for the Greedy method, together with the text: ``In Method A, the projects are selected simply based on the numbers of votes and total available budget. In each step, the method picks the most popular project that the remaining budget can fund. If the remaining available budget is not enough to fund the project, then the project is skipped. For example, the third most popular project in the following diagram is skipped as the remaining 50 CHF is not enough to fund the 60 CHF project. The process ends when no projects can be funded with the remaining budget.''}
		\label{fig:ex-greedy}
	\end{subfigure}
	\hfill 
	\begin{subfigure}{0.48\textwidth}
		\centering
		\includegraphics[width=\linewidth]{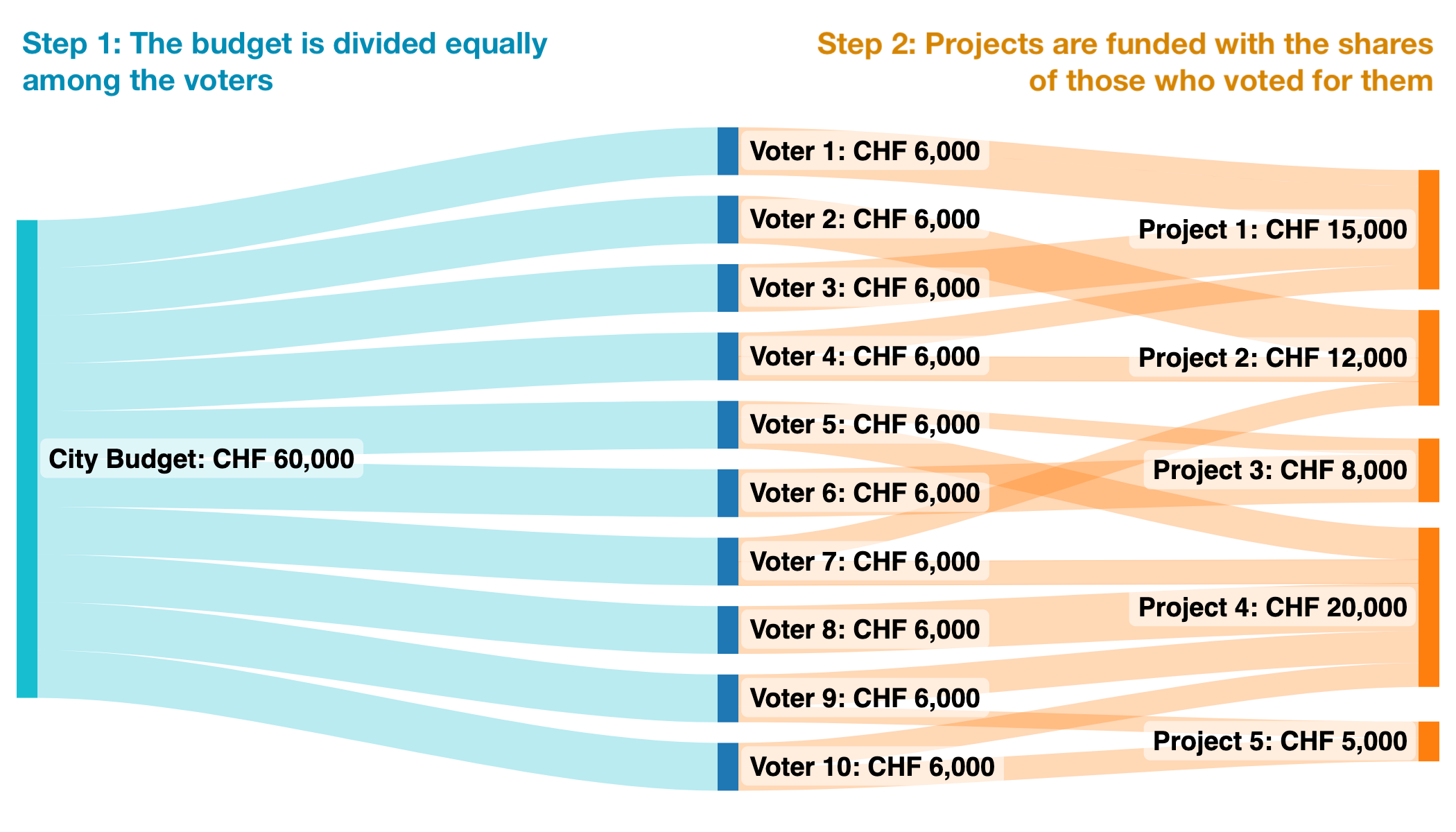}
		\caption{Illustration used for MES, together with the text: ``In the vote calculation of Method B, the total budget is divided so that each voter is assigned a hypothetical budget share to fund projects. The budget share assigned to a voter can only be used to fund projects that the voter has voted for. Method B goes through all project proposals, beginning with the projects with the highest number of votes. It selects a project if it can be funded using the budget shares of those who voted for the project. The method divides the cost of a project equally among its supporters.''}
		\label{fig:ex-mes}
	\end{subfigure}
	\caption{Illustrations used for the \textit{Mechanism Explanation Group} participants.}
	\label{fig:ex-mechanism}
\end{figure}
\begin{figure}[ht]
	\centering
	\begin{subfigure}{0.48\textwidth}
		\centering
		\includegraphics[width=0.95\linewidth]{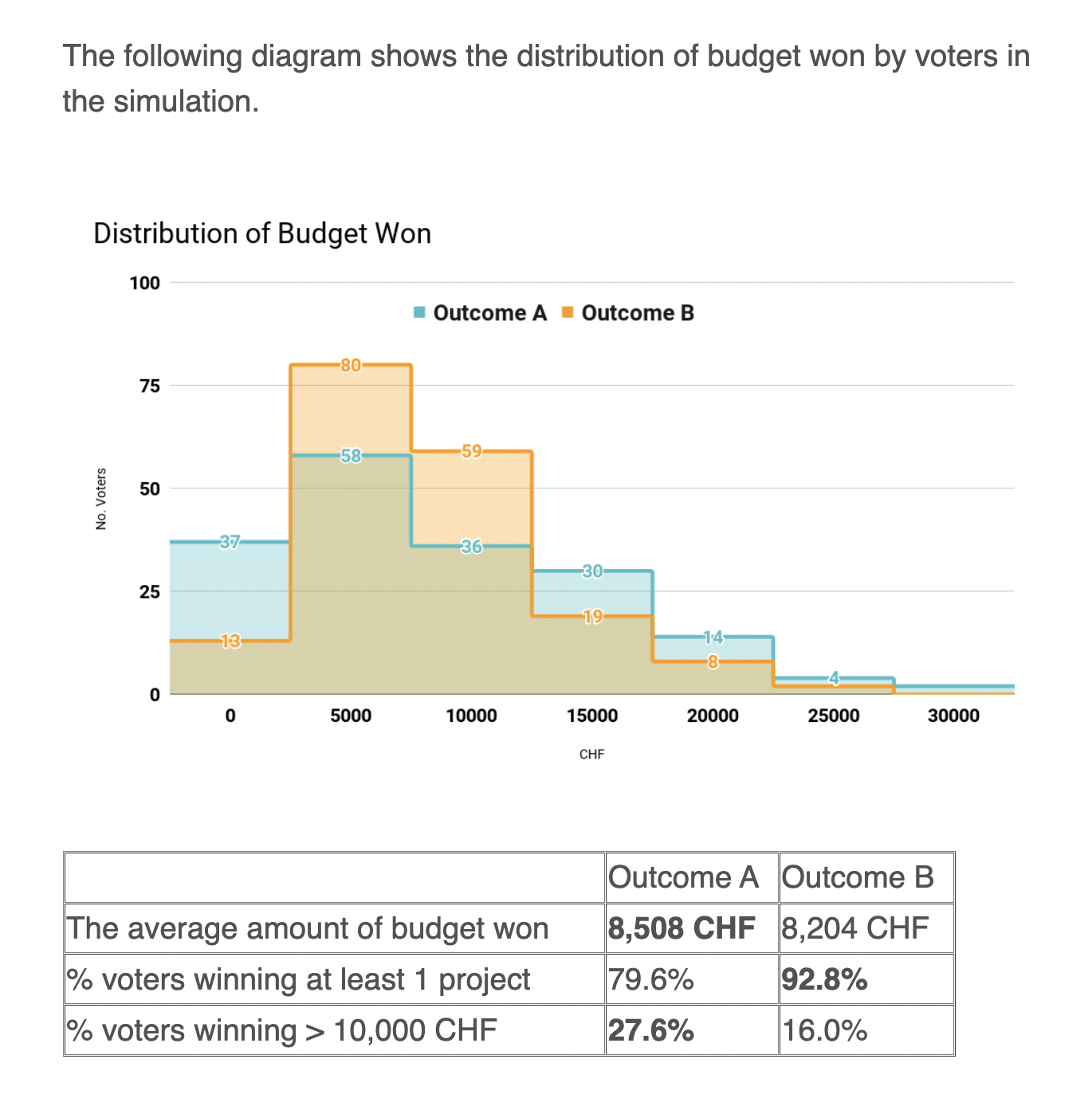}
		\caption{Version for Instance 1.}
		\label{fig:ex-indi-1}
	\end{subfigure}
	\hfill
	\begin{subfigure}{0.48\textwidth}
		\centering
		\includegraphics[width=0.95\linewidth]{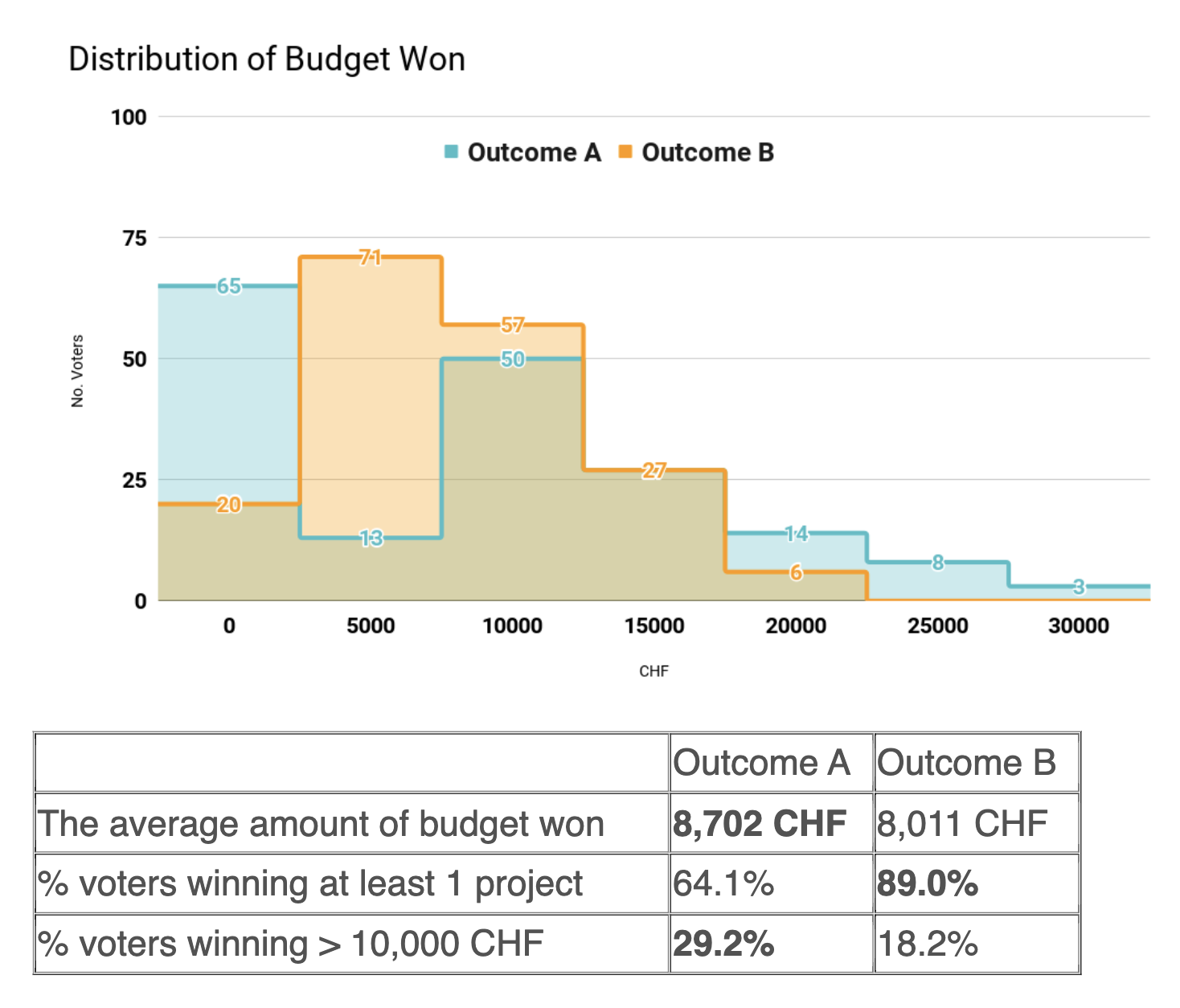}
		\caption{Version for Instance 2.}
		\label{fig:ex-indi-2}
	\end{subfigure}
	\caption{Illustration used for the \textit{Individual Explanation Group} (depending on the instance they were shown). The distribution diagram shows the number of voters who ``won'' a particular amount of budget, in the cost utility sense, together with a table offering related statistics.}
	\label{fig:ex-indi}
\end{figure}
\begin{figure}
	\begin{subfigure}{0.48\textwidth}
		\centering
		\includegraphics[width=0.95\linewidth]{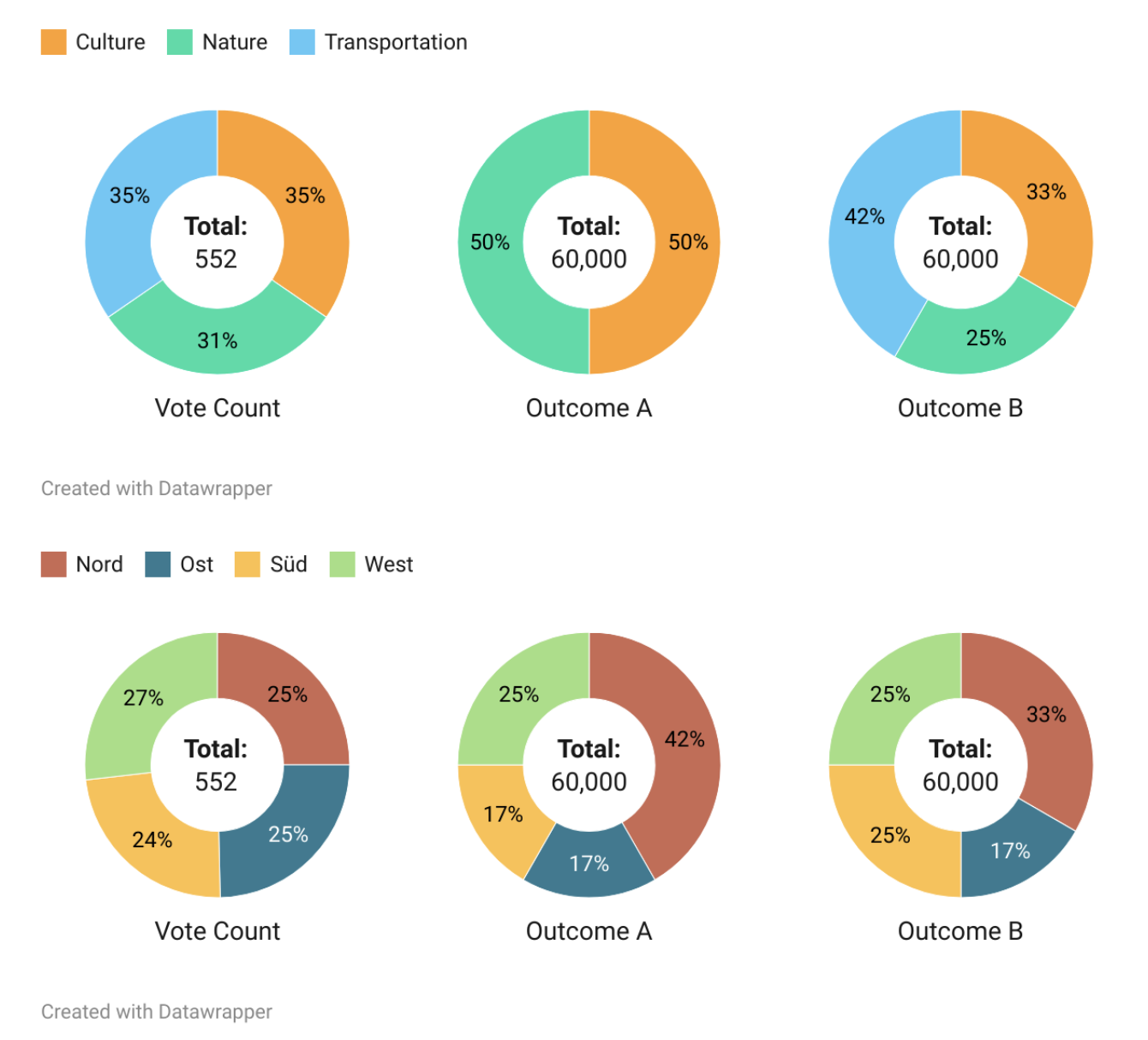}
		\caption{Version for Instance 1.}
		\label{fig:ex-group-1}
	\end{subfigure}
	\hfill 
	\begin{subfigure}{0.48\textwidth}
		\centering
		\includegraphics[width=0.95\linewidth]{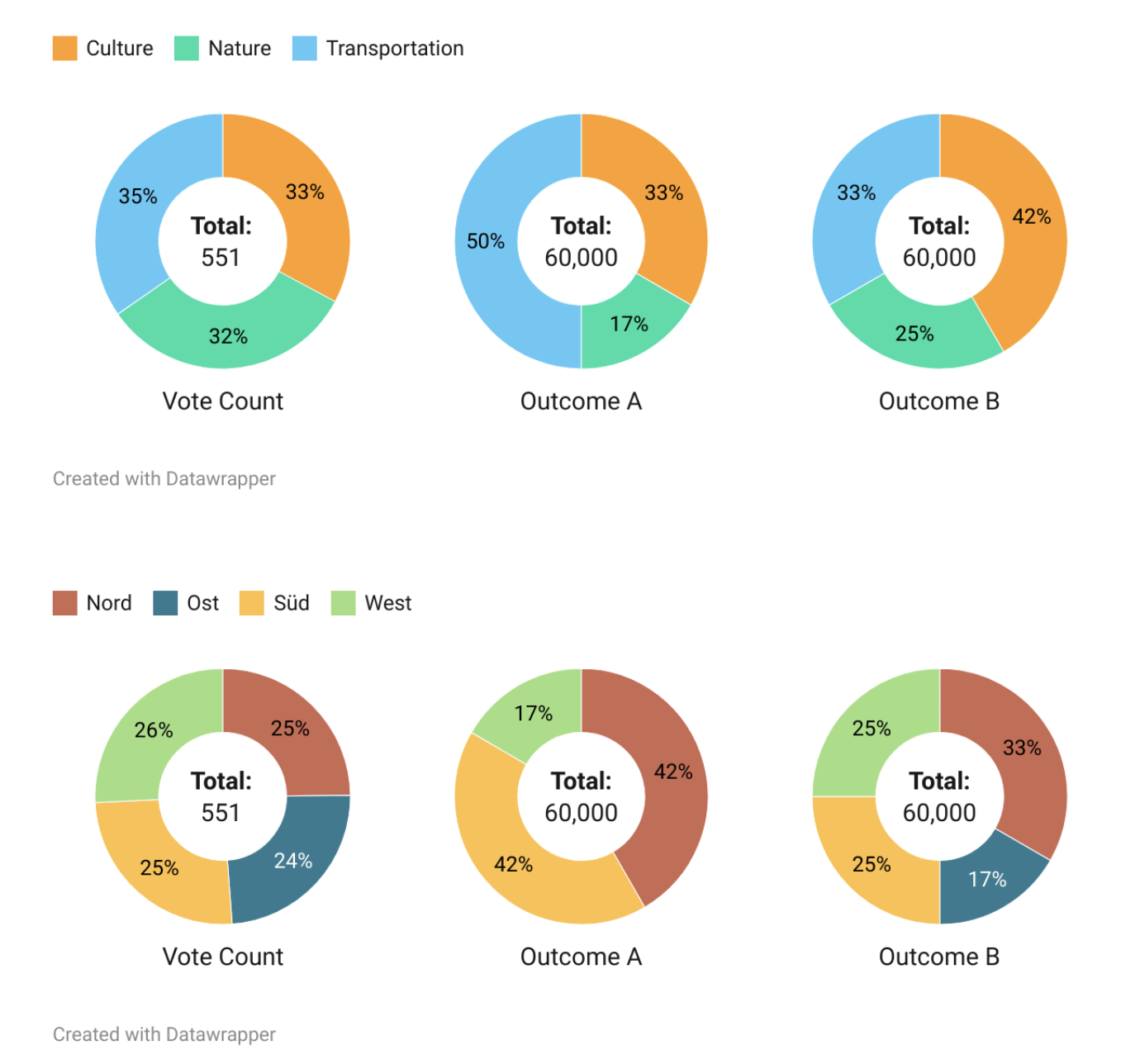}
		\caption{Version for Instance 2.}
		\label{fig:ex-group-2}
	\end{subfigure}
	\caption{Illustration used for the \textit{Group Explanation Group} (depending on the instance they were shown). The diagram shows the budget distributions across different districts and categories, offering insights into the allocated funds for each interest group in the simulated outcomes.}
	\label{fig:ex-group}
\end{figure}

\end{document}